 \documentclass[final,3p,times,twocolumn,a4paper]{elsarticle}
\usepackage[utf8]{inputenc}

\usepackage{graphicx}
\usepackage{amssymb}
\usepackage{amsmath}
\usepackage{tikz}
\usetikzlibrary{shapes.geometric,shapes.symbols}

\newcommand{\tikzsquare}[1]{\tikz[scale=0.3]{\fill[#1] (0,0) -- (0,1) -- (1, 1) -- (1, 0) -- cycle ;}}
\newcommand{\tikzrighttriangle}[1]{\tikz[scale=0.3]{\fill[#1] (0,0) -- (0,1) -- (1, 0.5) -- cycle ;}}

\newcommand{\tikzcircle}[1]{\tikz[scale=0.15]{\draw[fill=#1] circle(1) ;}}
\newcommand{\tikzstar}[1]{\tikz[scale=-1]{\node[scale=0.5,star,star points=5, star point ratio=2.25, fill=#1, draw] {};}}

\usepackage{lineno}

\usepackage{bm}% bold math
\usepackage{xcolor}
\usepackage{textcomp}
\usepackage{upgreek}
\usepackage{placeins}

\usepackage{setspace}

\journal{Journal of Non-Newtonian Fluid Mechanics}

\begin{document}

\begin{frontmatter}


\title{Viscometric Functions and Rheo-optical Properties of Dilute Polymer Solutions: Comparison of FENE-Fraenkel Dumbbells with Rodlike Models}% Force line breaks with \\

\author[chem]{I. Pincus}
\address[chem]{Department of Chemical Engineering, Monash University, 
Melbourne, VIC~3800, Australia}

\author[mcq]{A. Rodger}%
\address[mcq]{Molecular Sciences, Macquarie University, NSW, 2109, Australia}

\author[chem]{J. Ravi Prakash\corref{cor1}}
\ead{ravi.jagadeeshan@monash.edu}
\cortext[cor1]{Corresponding author. Tel +61 3 9905 3274; Fax +61 3 9905 5686.}

%\begin{spacing}{1.2}

\begin{abstract}
Rigid macromolecules or polymer chains with persistence length on the order of the contour length (or greater) have traditionally been modelled as rods or very stiff springs. 
The FENE-Fraenkel-spring dumbbell, which is finitely extensible about a non-zero natural length with tunable harmonic stiffness, is one such model which has previously been shown to reproduce bead-rod behaviour in the absence of hydrodynamic interactions. The force law for the FENE-Fraenkel spring reduces to the Hookean or FENE spring force law for appropriately chosen values of the spring parameters. It is consequently possible to explore the crossover region between the limits of bead-spring and bead-rod behaviour by varying the parameters suitably. In this study, using a semi-implicit predictor-corrector Brownian dynamics algorithm, the FENE-Fraenkel spring is shown to imitate a rod with hydrodynamic interactions when spring stiffness, extensibility and simulation timestep are chosen carefully. By relaxing the spring stiffness and extensibility, the FENE-Fraenkel spring can also reproduce spring-like behaviour, such as a crossover from $-1/3$ to $-2/3$ power-law scaling in the viscosity with shear rate, and a change from positive to negative second normal stress difference.
Furthermore, comparisons with experimental data on the viscosity and linear dichroism of high aspect ratio, rigid macromolecules shows that the extensibility and stiffness of the FENE-Fraenkel spring allows for equal or improved accuracy in modelling inflexible molecules compared to rodlike models. 
\end{abstract}


%\end{spacing}

\end{frontmatter}

%%%%%%%%%%%%%%%%%%%%%%%%%%%%%%%%%%%%%%%%%%%%%%%%%%%%%%%%%%%%%%%%%%%%%%%%%%%%%%%%%%%%%%%%%%%%%%%%%%%%%%%%%%%%%%%%%%%%%%%%%%%%%%%%%%%%%%%%%%%%%%%%%%%%%%%%%%%%%%%%%%%%%%%%%%%%%%%%%%%%%%%%%%%%%%%%%%%%%%%%

\section{Introduction}

While both bead-rod and bead-spring models are able to successfully predict certain rheological properties of polymer chains in dilute solution, their relative usefulness continues to be debated.
This debate began with the theoretical finding by Kramers \cite{kramers1944viscosity} that the equilibrium distribution of the included angle of a trimer differs depending on whether the monomer links are represented as constrained rods or infinitely stiff springs (the reason for which is discussed by Van Kampen \cite{van1981statistical} and Lodder \textit{et al.} \cite{VanKampen1984}). 
It is often argued that true constraints do not exist in nature and therefore stiff springs should be preferred, but in fact the form of the spring potential can also affect the included angle, such that a quantum mechanical treatment appears to be necessary for a fully correct solution given a real polymer \cite{rallison1979role, van1981statistical}. 
Nevertheless, the practical difference is small, as both models yield results in qualitative agreement with experimental studies \cite{Shaqfeh2005, Larson2005}, while direct quantitative comparisons are sparse for bead-rod chains \cite{Prakash2019, pan2018shear}.

Some of the most stark differences between predictions of bead-rod and bead-spring models arise for dilute polymer solutions in shear flow, as described by Pan et al. \cite{pan2018shear}.
For example, there is no clear consensus on the shear-thinning exponent of the viscosity with shear rate for high molecular weight polymers. Some experimental studies using viscometers show a $-(2/3)$ power law scaling in the viscosity of high molecular weight flexible polymers under shear flow, followed by a high-shear Newtonian viscosity plateau  \cite{Hua2006}. 
Other studies of single fluorescently stained DNA molecules show a power law decay of viscosity with shear rate, with a shear thinning exponent around $-0.54$~\cite{teixeira2005shear, schroeder2005dynamics}. Confusingly, bead-spring and bead-rod models each show different aspects of this experimental behaviour, depending on whether effects such as finite extensibility, chain flexibility, hydrodynamic interactions or excluded volume are included. 

Early calculations by Stewart and Sorensen \cite{Stewart:1972gt} for a rigid dumbbell with hydrodynamic interactions included found an eventual $-(1/3)$ power law scaling in the viscosity with shear rate after an initial Newtonian plateau. 
Using a similar method for FENE dumbbells without hydrodynamic interactions, Fan \cite{Fan:1985jk} found a shear-thinning exponent of $-0.607$, while the analytical pre-averaged FENE-P dumbbell model gives an asymptotic exponent of $-(2/3)$ \cite{bird1987dynamics}. 
These dumbbell models cannot reproduce the high-shear second Newtonian plateau. 
For moderate-length (20 to 100 bead) bead-rod chains without hydrodynamic interactions, an exponent of $-(1/2)$ was found\cite{Liu1989, doyle1997dynamic} (which appears to change to $-0.3$ as hydrodynamic interactions are included \cite{Petera1999}), along with a second Newtonian plateau at high shear. However, in the presence of both hydrodynamic and excluded volume interactions, Petera and Muthukumar found no high-shear Newtonian viscosity plateau \cite{Petera1999}. 
For finitely extensible bead-spring chains with hydrodynamic interactions and excluded volume, there is an intermediate $-(1/2)$ power law regime, followed by a $-(2/3)$ \cite{jendrejack2002stochastic} or $-0.61$ \cite{schroeder2005dynamics} exponent at very high shear rates. 
Clearly, although both the bead-spring and bead-rod models purport to represent the physics of the same underlying polymer, there are fundamental differences in the high-shear viscosity scaling.

Additionally, rods exhibit an instantaneous stress jump at the inception of flow (which is also found in experimental studies \cite{Mackay1992}) arising from the viscous fourth-order contribution to the stress tensor~\cite{bird1987dynamics, doyle1998dynamic, doyle1997dynamic}. Bead-spring models do not contain this term for any form of the spring force law, so that additional physics in the form of a parallel dissipative dashpot is needed to produce an instantaneous stress jump \cite{manke1992stress, gerhardt1994relationships, kailasham2018rheological}. 

In general, it is unclear whether these differences between bead-rod and bead-spring models arise due to the form of the spring force law, the effects of rigid constraints, or additional physics such as excluded volume or hydrodynamic interactions \cite{pan2018shear}. 
These differences are difficult to investigate systematically, as there is currently no model which can represent the full range of possible parameters.
For example, the FENE and Marko-Siggia WLC force laws are used to coarse-grain many Kuhn lengths of a polymer chain into a single bead-spring segment, and so are physically unrealistic or unusable as representations of short, rigid, inextensible sections of chain.
While spring force laws have been developed which can be used at the level of a single Kuhn segment \cite{underhill2004coarse, Saadat2016}, they go smoothly to zero force at low extensions, rather than having a `negative' force which opposes compression and keeps the spring length constrained as for a rod.
A model with a non-zero natural length, such as the Fraenkel force law, is able to approximate a rigid chain segment in the high-stiffness limit, but its strictly linear behaviour means it cannot model a finitely extensible section of chain.
By developing a spring force law which can represent both a stiff, inextensible rod, as well as a finitely extensible entropic spring, it may be possible to directly examine many of the observed differences which arise when using either a bead-rod or bead-spring model of a flexible polymer molecule.

\begin{figure*}[t]
    \centerline{
    \begin{tabular}{c c}
        \includegraphics[width=80mm]{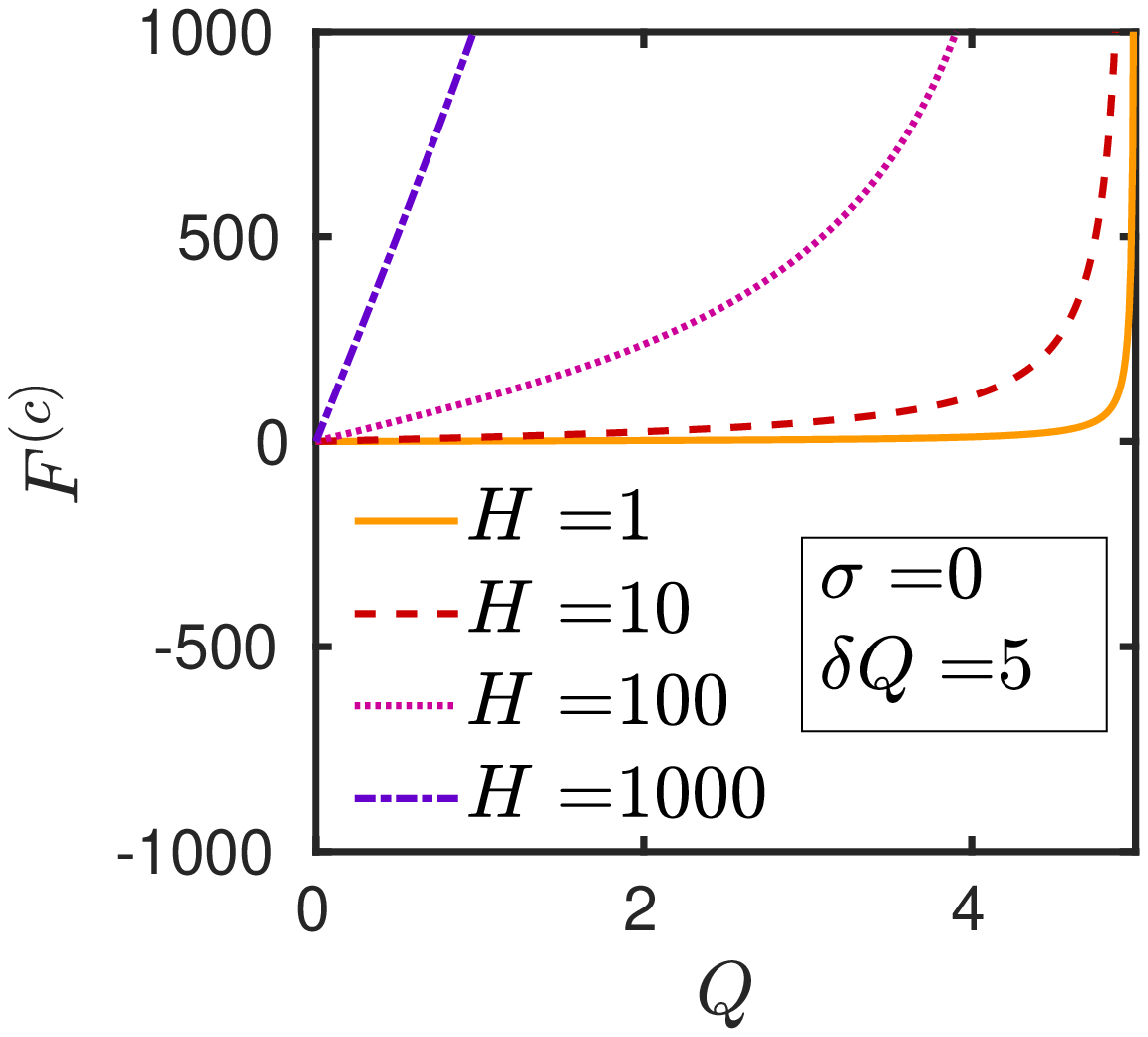} 
        & \includegraphics[width=80mm]{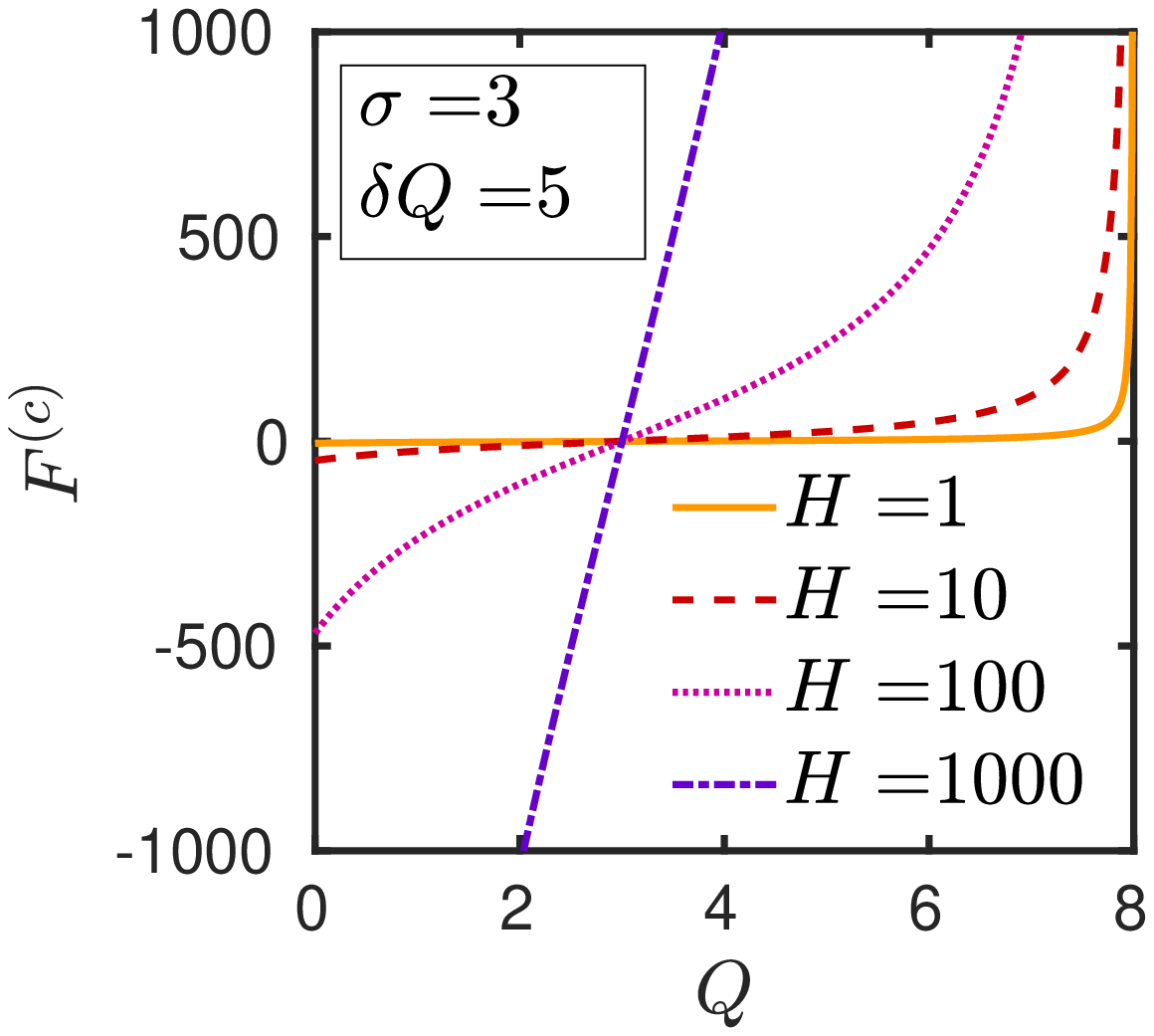} \\
        (a) & (b) \\
        \includegraphics[width=80mm]{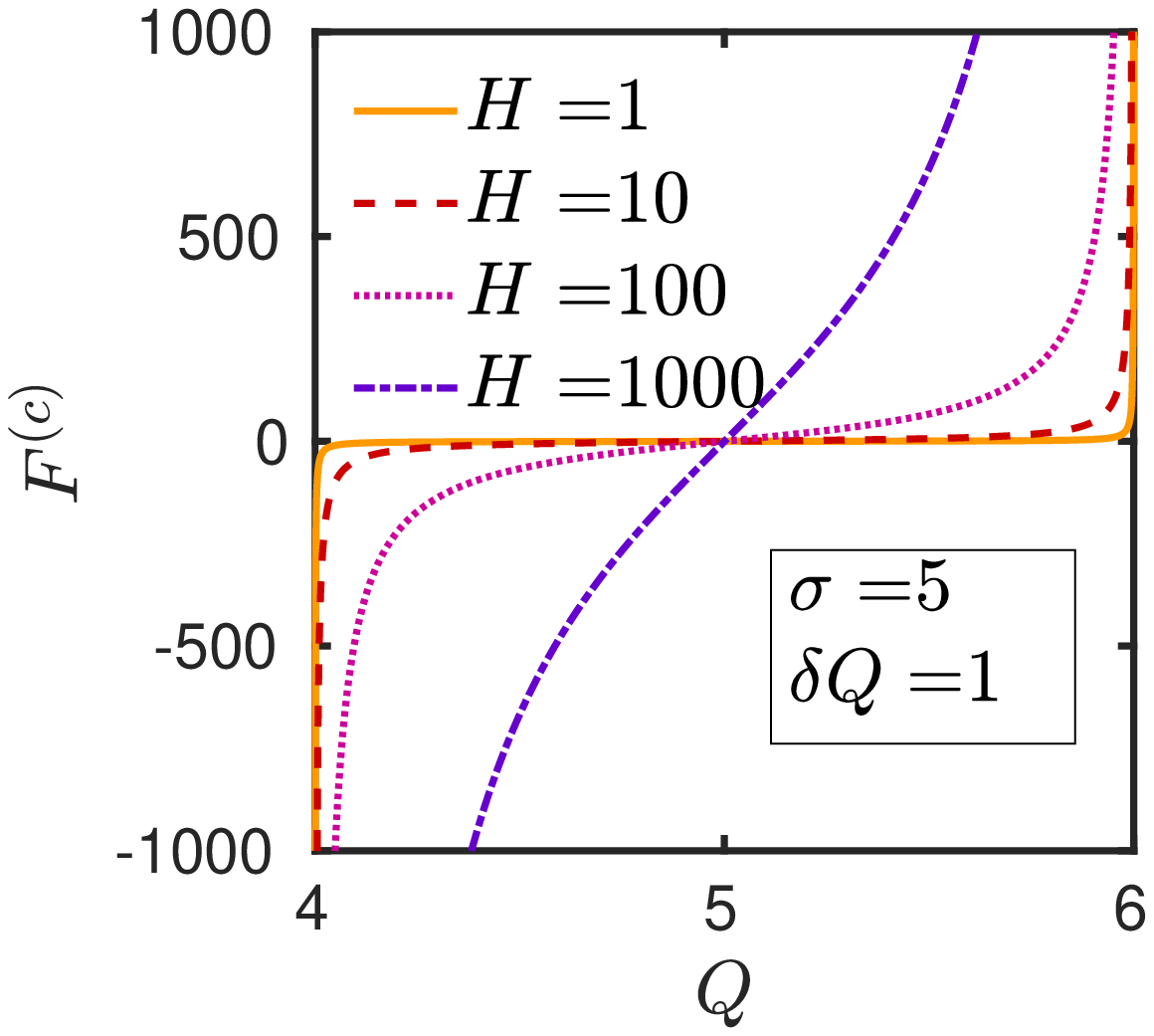} 
        & \includegraphics[width=80mm]{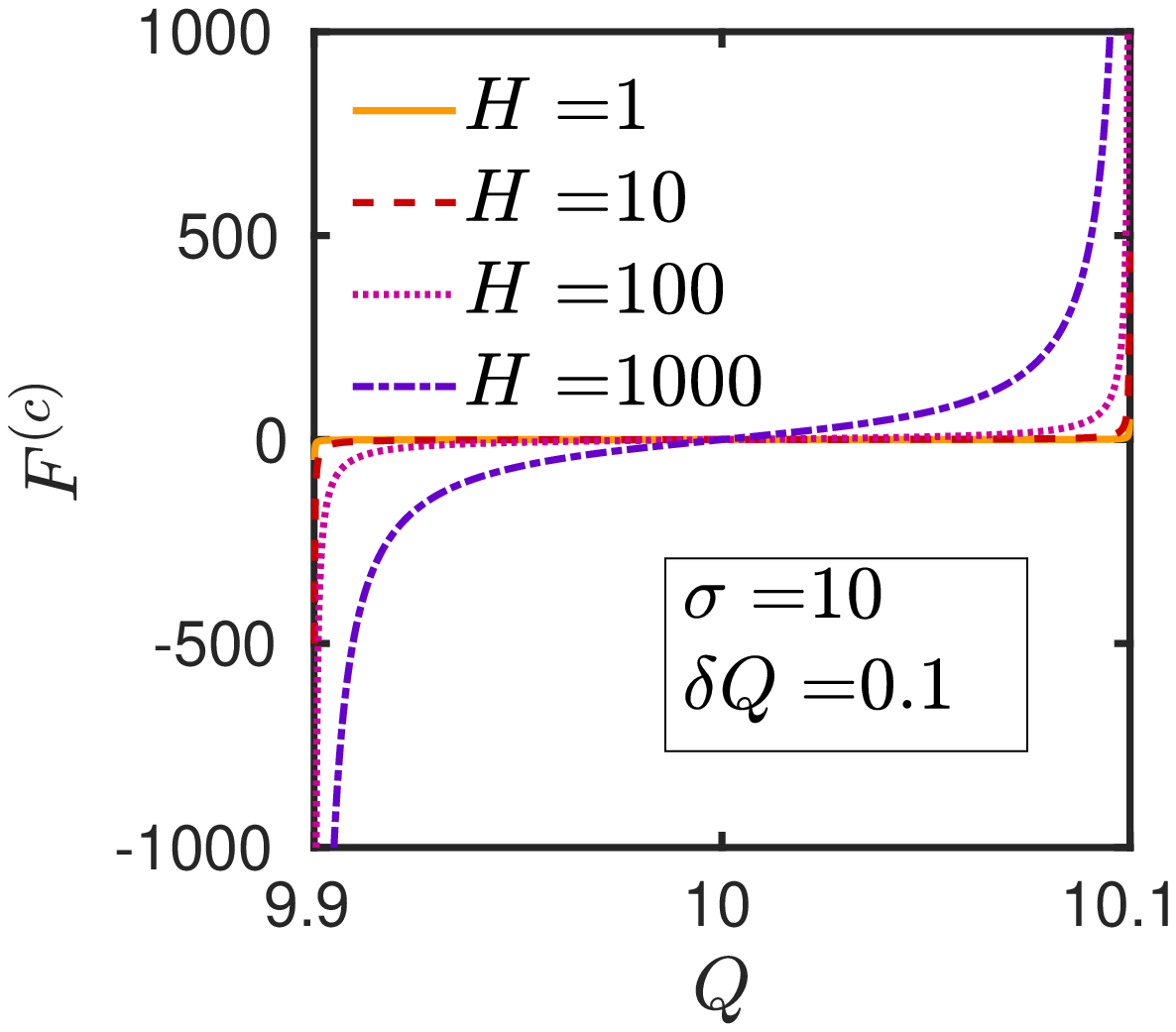} \\
        (c) & (d) \\
    \end{tabular}
    }
 \vskip-5pt   
    \caption[Visualisation of the spring force for a FENE-Fraenkel spring]{Visualisation of the spring force for a FENE-Fraenkel spring with varying values of spring stiffness $H$, natural length $\sigma$ and extensibility $\delta Q$. The force is linear around $Q=\sigma$, but quickly approaches $\pm \infty$ as $Q\rightarrow Q\pm \delta Q$. (a) Corresponds to the limit of a FENE spring. (b) $\delta Q > \sigma > 0$, leading to a negative, but not infinite force for $Q < \sigma$. (c) Approximate limit of a Fraenkel spring for large $\delta Q$. (d) Large $\sigma$ and small $\delta Q$ leads to tightly constrained $Q$ and hence approximation of inextensible rod.}
    \label{FF_force}
\end{figure*}

The so-called FENE-Fraenkel spring force law \cite{Hsieh2006} may be viable as a way to model both rods and entropic springs, as well as the crossover between the two. 
For this spring, the force between connected beads at a given extension Q (where the bead-bead vector is denoted by $\bm{Q}$) is
\begin{equation}
    \bm{F}^{(c)} = \frac{H(Q-\sigma)}{1-(Q-\sigma)^2/(\delta Q)^2} \frac{\bm{Q}}{Q}
\label{FF_force_eqn}
\end{equation}
where $\bm{F}^{(c)}$ is the force vector between the beads, $\sigma$ is the natural length of the spring ($Q=\sigma$ in the absence of any additional forces), $\delta Q$ is the maximum extensibility around $\sigma$, and $H$ is the spring stiffness. 
By examining Fig.~\ref{FF_force}, we can see that the spring length $(Q)$ will never extend more than a distance $\pm \delta Q$ from the `natural' length $Q=\sigma$. 
In addition, the spring is approximately linear  close to $Q=\sigma$. 
Note that we can recover the FENE force law by setting $\sigma=0$, in which case we can identify that $\delta Q \equiv Q_0$, where $Q_0$ is the maxmimum extensibility of the FENE spring, as displayed in Fig.~\ref{FF_force}~(a).
When $\sigma$ is further increased such that $\delta Q > \sigma > 0$, the spring behaves similarly to a FENE spring at large extensions but has a non-zero rest length and a negative, but not divergent force at $Q< \sigma$, as seen in Fig.~\ref{FF_force}~(b). 
Furthermore, in the limit that $\delta Q \rightarrow \infty$, we recover the Fraenkel spring, as can be seen approximately in Fig.~\ref{FF_force}~(c).
Finally, setting $\delta Q \rightarrow \infty$ and $\sigma = 0$ gives the simple Hookean spring.
In order to simulate a rod, a large $\sigma$ and small $\delta Q$ can be used as in Fig.~\ref{FF_force}~(d), such that the spring length is constrained to lie between the two limits with a divergent force at $Q = \sigma \pm \delta Q$, leading to an approximation of an inextensible rod.

The FENE-Fraenkel spring was first introduced by Larson and coworkers \cite{Hsieh2006} as a way to mimic a freely jointed bead-rod chain without the significant computational complexities associated with constrained Brownian dynamics simulations necessitated by the use of a rod as connector between beads.
They were able to show that the free-draining material properties of a FENE-Fraenkel spring chain match those of a bead-rod chain in shear and extensional flow with reduced CPU time when the spring parameters are chosen appropriately, namely with sufficiently high $H$ and low $\delta Q$. 
However, when hydrodynamic interactions were included, there were differences in the shear rate dependent viscosity between the bead-rod chain and bead-FENE-Fraenkel spring chain \cite{Larson2005}.
Since there are no analytical results for rods or springs with this model and flow geometry, it was unclear whether the differences were due to simulation artefacts, or whether the differences were intrinsic to the models. 
Additionally, they only used the FENE-Fraenkel spring as a way to represent a rod, and did not explicitly explore the crossover between a spring and a rod.

In the current work, the properties of a simple dumbbell connected via a FENE-Fraenkel spring will be examined, which can be used to address several of the open questions mentioned above. 
Specifically, we investigate whether a stiff and inextensible FENE-Fraenkel spring can be used as a replacement for a rod in shear flow in terms of material property scaling with shear rate, the stress jump, and failure to adhere to the stress-optical law.
Since the full distribution function of a bead-rod dumbbell with HI can be determined semi-analytically in shear flow \cite{Stewart:1972gt, bird1987dynamics}, simulations can be compared with exact results.
Once similarities between bead-spring and bead-rod dumbbells are established, the differences induced by altering spring stiffness and extensibility can be systematically examined, eventually converging on the FENE, Fraenkel and Hookean spring limits.

There are also a variety of experimental results for the viscosity of rigid molecules with high aspect ratios, such as poly($\gamma$-benzyl L-glutamate) \cite{Mead1990} and tobacco mosaic virus \cite{Wada1954}, for which the rigid-rod model has been shown to be qualitatively and reasonably quantitatively accurate (for example, see Fig. 14.4 in \cite{bird1987dynamics} or Fig. 5 in \cite{yang1958non}). 
These filamentous molecules with contour lengths on the order of persistence length have traditionally been modelled as rods \cite{Dhont2007}, while they do in fact exhibit some flexibility and finite extensibility about their equilibrium end-to-end distance. 
The form of the FENE-Fraenkel spring force law allows for investigation of the independent effects of spring stiffness, natural length and extensibility through variation of $H$, $\sigma$ and $\delta Q$, such that it may be a better qualitative model of the extensibility of these molecules compared to rigid rod models. 
However, note that the FENE-Fraenkel spring is not being developed as an actual physical model of a length of semiflexible polymer chain, as the form of the FENE-Fraenkel force law cannot accurately reproduce either the high-extension or high-compression behaviour of a short section of a wormlike chain \cite{Yamakawa2016}, and chain bending is not accounted for.

Besides viscometric functions and rheo-optical properties derived from the stress and gyration tensors, the predicted linear dichroism (LD) of the dumbbell ensemble is computed. 
Linear dichroism refers to the difference in absorption of parallel and perpendicularly polarised light by an oriented ensemble of molecules, and can be used to investigate the structure and interaction of large, flexible macromolecules which are hard to characterise using traditional techniques such as crystallography or NMR. 
As an analytic technique, LD may be useful for high-throughput screening of potential drug targets to DNA or cytoskeletal proteins \cite{Rodger2009}, however difficulty in predicting the orientation of large, flexible molecules inhibits its quantitative accuracy \cite{McLachlan2013}. 
Since the LD signal is related to molecular orientation, it provides a useful experimental test of models of polymer behaviour in flow, and these models may in turn be crucial for future progress in improving the LD technique. 
To our knowledge, this paper represents the first direct comparison of a BD simulation with experimental LD data, although BD appears to have been used both to compute orientation of semiflexible chains for comparison with LD data \cite{OdegaardJensen1996}, and also to interpret previous experimental data for LD of biomolecules in lipid membranes \cite{castanho2003using}.
Additionally, the semi-analytical rod models used in this paper have previously been applied to rigid bacteriophage LD by McLachlan et al. \cite{McLachlan2013}.

This article is split into 3 further sections. In section 2, we set up the theoretical treatment of both bead-spring and bead-rod dumbbells which will inform simulations and describe our methods for numerically evaluating the distribution functions.
In section 3, we discuss our simulation results, particularly highlighting the comparison between bead-rod and bead-spring dumbbell material functions. 
Finally, we conclude with the key findings of our work and future plans for bead-FENE-Fraenkel-spring-chain simulations. 

%%%%%%%%%%%%%%%%%%%%%%%%%%%%%%%%%%%%%%%%%%%%%%%%%%%%%%%%%%%%%%%%%%%%%%%%%%%%%%%%%%%%%%%%%%%%%%%%%%%%%%%%%%%%%%%%%%%%%%%%%%%%%%%%%%%%%%%%%%%%%%%%%%%%%%%%%%%%%%%%%%%%%%%%%%%%%%%%%%%%%%%%%%%%%%%%%%%%%%%%

\section{Methods}

\subsection{Governing Equations for Dumbbell Models}
\label{governing equations}

The general dumbbell model consists of two massless beads of radius $a$, connected by either a spring, or a rod with length $L$. 
We describe the co-ordinates of the two beads by $\bm{r}_1$ and $\bm{r}_2$, with the connector vector between the beads given by $\bm{Q} = \bm{r}_2 -\bm{r}_1$. For a rod, this vector can be simplified as $\bm{Q} = L \bm{u}$, where $\bm{u}$ is the radial unit vector in spherical coordinates. For a spring, the unit vector in the direction of the spring is $\bm{u} = \bm{Q}/Q$, where $Q$ is the dummbell length.

The dumbbell is suspended in an incompressible Newtonian solvent of viscosity $\eta_s$, with a velocity field imposed by shear flow expressed in the form:
\begin{equation}
\label{velocity field}
    \bm{v}(\bm{r}, t) = \bm{v}_0(t) + \bm{\kappa}(t)\cdot \bm{r}
\end{equation}
Here $\bm{r}$ is a position given with respect to a fixed reference frame (the laboratory frame), $\bm{v}_0$ is a position-invariant vector and $\bm{\kappa}(t)$ is a tensor given by the transpose of the velocity field gradient, which is also position-invariant. 
Both $\bm{v}_0$ and $\bm{\kappa}$ can in general be a function of time.

If we include hydrodynamic interactions in the form of a diffusion tensor $\bm{\Omega}$ (which we have chosen as the Rotne-Prager-Yamakawa tensor), then the differential equation for the time evolution of the bead-spring dumbbell connector vector distribution function $\psi = \psi(t,\bm{Q})$ is given by the following Fokker-Planck equation \cite{bird1987dynamics}:
\begin{multline}
\label{Fokker-Planck Hookean Units}
    \frac{\partial \psi^*_\text{H}}{\partial t^*_\text{H}} = -\frac{\partial}{\partial \bm{Q}^*_\text{H}} \cdot \left\{ \bm{\kappa}^*_\text{H}\cdot \bm{Q}^*_\text{H} - (\bm{\delta}-\zeta \bm{\Omega}) \cdot \frac{1}{2} \bm{F}^{*(c)}_\text{H} \right\} \psi^*_\text{H} \\
    + \frac{1}{2} \frac{\partial}{\partial \bm{Q}^*_\text{H}} \frac{\partial}{\partial \bm{Q}^*_\text{H}} \bm{:} [\bm{\delta}-\zeta \bm{\Omega}] \psi^*_\text{H}
\end{multline}
where $\bm{\delta}$ denotes the unit tensor and length, time and force variables have been re-scaled in terms of `Hookean' units respectively, such that
\begin{equation}
    l_\text{H} \equiv \sqrt{\frac{k_\text{B} T}{H}}, \lambda_\text{H} \equiv \frac{\zeta}{4H}, F_\text{H} \equiv \sqrt{k_\text{B} T H}
\label{Hookean_system}
\end{equation}
and non-dimensional variables are denoted with a star superscript, such as $Q_\text{H}^* = Q/l_\text{H}$ or $t^*_\text{H} = t/\lambda_\text{H}$. 
The friction coefficient $\zeta$ is equivalent to that for a sphere in Stokes flow, so that $\zeta = 6\pi \eta_s a$. 
It is also possible to use another system for non-dimensionalising our simulations and results, which we denote the `rodlike' unit system and identify:
\begin{equation}
    l_\text{R} \equiv \sigma, \lambda_{R} \equiv \frac{\sigma^2 \zeta}{k_\text{B} T}, F_{R} \equiv \frac{k_\text{B} T}{\sigma}
\label{Rodlike_definitions}
\end{equation}
Note that in this unit system, the rodlike FENE-Fraenkel spring stiffness is given by $H^*_R = H\sigma^2 /k_\text{B} T$. 
This system is commonly used for bead-rod models, with rod length $L$ instead of natural length $\sigma$ in Eq.~\eqref{Rodlike_definitions} above. 
Since the times, length and forces are scaled in the same way between FENE-Fraenkel springs and rods in this unit system, we can compare results directly between these two models without having to convert back to dimensional form. 

It is fairly straightforward to convert between the `rodlike' and `Hookean' unit systems via substitution, for example for lengths we have
\begin{equation}
\begin{aligned}
\label{QH to QR conv}
    Q^*_\text{H} &  = \frac{Q}{l_\text{H}} = \frac{Q}{\sqrt{k_\text{B} T / H}} \\
    & = Q^*_\text{R} \sigma \cdot \sqrt{\frac{k_\text{B} T H^*_\text{R}}{k_\text{B} T \sigma^2}}  =  Q^*_\text{R} \sqrt{H^*_\text{R}}
\end{aligned}       
\end{equation}
\begin{equation}
    Q^*_\text{R} = \frac{Q}{\sigma} = \frac{Q^*_\text{H} \sqrt{k_\text{B} T / H}}{\sigma^*_\text{H} \sqrt{k_\text{B} T / H}} = \frac{Q^*_\text{H}}{\sigma^*_\text{H}} 
\end{equation}
and for times we have
\begin{subequations}
\label{th to tr conv}
\begin{equation}
    t^*_H = \frac{t}{\lambda_H} = t^*_R \frac{4H}{\zeta} \frac{\sigma^2 \zeta}{k_\mathrm{B} T} = 4 H^*_R t^*_R
\end{equation}
\begin{equation}
    t^*_R = \frac{t}{\lambda_R} = t^*_H \frac{\zeta}{4H} \frac{k_\mathrm{B} T}{\sigma^2 \zeta} = \frac{1}{4{\sigma^*_H}^2} t^*_H 
\end{equation}
\end{subequations}

Finally, we also define rodlike and Hookean hydrodynamic interaction parameters $h^*_H$ and $h^*_R$, which are essentially dimensionless bead radii, as:
\begin{equation}
    h^*_H = \frac{a}{\sqrt{\pi} l_H} \equiv \frac{a^*_H}{\sqrt{\pi}}
\end{equation}
\begin{equation}
    h^*_R = \frac{3 a}{4 \sigma} \equiv \frac{3 a^*_R}{4}
\end{equation}
Note that this implies that $h^*_H = (4 \sqrt{H^*_R})/(3 \sqrt{\pi}) h^*_R$, or equivalently that $h^*_R = (3 \sqrt{\pi})/(4 \sigma^*_H) h^*_H$.

The Fokker-Planck Eq.~\eqref{Fokker-Planck Hookean Units} can be expressed as an equivalent stochastic differential equation via It\^{o}'s calculus \cite{Ottinger1996}, which is then integrated over thousands to millions of trajectories using a semi-implicit predictor-corrector scheme. This numerical scheme has been detailed by other authors for both FENE \cite{somasi2002brownian, Hsieh2003, Prabhakar2004} as well as FENE-Fraenkel \cite{Hsieh2006} springs. A full description of the derivation of distribution functions and numerical integration procedures used here for both the FENE-Fraenkel bead-spring dumbbell and rodlike models can be found in the supporting information sections 1 and 2. A given FENE-Fraenkel dumbbell simulation for a single trajectory therefore requires specification of two dimensionless spring parameters, either $\sigma^*_H$ and $\delta Q^*_H$ for Hookean units or $H^*_R$ and $\delta Q^*_R$ for rodlike units; the timestep $\Delta t^*_H$ or $\Delta t^*_R$, with the conversion between the two influenced by the choice of spring parameters as per Eq.~\eqref{th to tr conv}; the flow field $\bm{\kappa}^*_H$ or $\bm{\kappa}^*_R$, which varies with inverse time and therefore follows similar scaling to that in Eq.~\eqref{th to tr conv}; and finally the hydrodynamic interaction parameter $h^*_R$ or $h^*_H$, which again are proportional to each other based on the spring parameters as per Eq.~\eqref{hstar dimensionalisation}. To ensure sufficient sampling for low error, the ensemble of dumbbells was allowed to reach steady-state, and then data was collected for tens to hundreds of relaxation times. Ensemble size was generally of order $10^6$ to $10^8$ dumbbells.

The rodlike distribution function is solved semi-analytically via a harmonic expansion, as was originally done by Stewart and Sorensen \cite{Stewart:1972gt}. 
This method was extended to include RPY HI between beads as per Bird \textit{et al.} \cite{bird1987dynamics}, and to solve for the transient distribution function as detailed by McLachlan and coworkers \cite{McLachlan2013}.
Note that the same solution method can be used for a variety of rodlike models which have the same general form of the diffusion equation, such as multibead-rods, prolate spheroids and slender bodies. 
A full description of the semi-analytical solution method used here can be found in the supporting information, section 2.1.

\subsection{Measured Quantities}

Viscometric properties of the solution are calculated using the Kramers form of the stress tensor \cite{bird1987dynamics}. For springs, this tensor has the form
\begin{equation}
\label{Bead-Spring-Dumbbell_stress_tensor}
    \bm{\tau}_p = \bm{\tau} + \eta_s \bm{\dot{\gamma}} = -n \langle \bm{Q} \bm{F} \rangle + n k_\text{B} T \bm{\delta}
\end{equation}
where $\bm{\delta}$ is the unit tensor, $\eta_s$ is the solvent viscosity and $\bm{\tau}_p$ is the polymer contribution to the stress tensor. 
For rods, the stress tensor is given by
\begin{equation}
\label{Bead-Rod-Dumbbell_stress_tensor_with_HI}
    \bm{\tau}_p = - 3nkT\langle \bm{u u} \rangle - nkT\lambda \{\bm{\kappa} \bm{:} \langle \bm{u u u u} \rangle \}+ nkT \bm{\delta}
\end{equation}
where $\lambda$ is a time constant which varies with the specific rod model and form of HI \cite{bird1987dynamics}. 
The viscosity $\eta_p$, first and second normal stress differences $\Psi_{1}$ and $\Psi_{2}$ and an angle $\chi_\tau$ characterising the tensor orientation can be calculated using this stress tensor. In shear flow, these quantities have the form:
\begin{subequations}
\begin{equation}
    -\eta_p = \frac{\tau_{xy}}{\dot{\gamma}}
\end{equation}
\begin{equation}
    -\Psi_1 = \frac{\tau_{xx} - \tau_{yy}}{\dot{\gamma}^2}
\end{equation}
\begin{equation}
    -\Psi_2 = \frac{\tau_{yy} - \tau_{zz}}{\dot{\gamma}^2}
\end{equation}
\end{subequations}
\begin{equation}
\label{chi tau definition}
    \chi_\tau = \frac{1}{2}\arctan{\frac{2 \tau_{xy}}{\tau_{xx} - \tau_{yy}}} = \frac{1}{2}\arctan{\frac{2 \eta_p}{\Psi_1 \dot{\gamma}}}
\end{equation}
A similar characteristic angle $\chi_G$ can be found for the second moment $\langle \bm{QQ} \rangle$:
\begin{equation}
    \chi_G = \frac{1}{2}\arctan{\frac{2 \langle Q_{xy}\rangle}{\langle Q_{xx} - Q_{yy}\rangle}}
\end{equation}
where $Q_{mn} \equiv Q_{m} Q_{n}$. 

While $\chi_G$ and $\chi_{\tau}$ are equal at equilibrium, they may differ in the presence of flow.
Since the $\chi_G$ parameter is related to the measured optical birefringence through the refractive index tensor \cite{janeschitz2012polymer, fuller1995optical}, while the $\chi_\tau$ parameter gives the orientation of the stress tensor, when $\chi_G$ and $\chi_\tau$ are not equal, the stress optical rule will not hold.
The stress optical rule is generally found to hold for small extensions of high molecular weight, highly extensible polymers, but does not hold for dilute solutions of rigid rods \cite{fuller1995optical}.

In addition to these viscometric functions, the predicted orientation parameter or `$S$-parameter' of the ensemble will also be measured. For shear flow along the $x$-axis, the $S$-parameter is defined as \cite{Norden1978, McLachlan2013}:
\begin{equation}
\label{$S$-parameter expression}
    S = \frac{1}{2}\Big[3\langle \cos^2{\theta_s} \rangle - 1\Big]
\end{equation}
where $\theta_s$ is the angle between the flow direction and the molecular orientation axis, such that in this case $\cos {\theta_s} = \bm{\delta}_x \cdot \bm{u}$ where $\bm{\delta}_x$ is the unit vector in the $x$-direction.

The $S$-parameter is proportional to the measured Linear Dichroism (LD) of a sample.
The LD of a sample arises from the transition dipole moment of its constituent molecules, which will absorb light polarised parallel to the moment vector but does not interact with light polarised perpendicularly \cite{Norden1978, Rodger2009}. 
\begin{figure}[pthb]
  \centering
  \includegraphics[width=8.5cm,height=!]{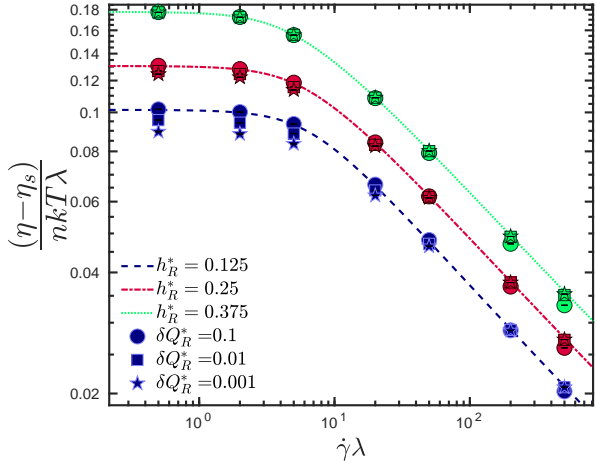}
  \vskip-2pt 
  \caption[Numerical artefacts in FENE-Fraenkel spring simulations for viscosity measurements]{Comparison of FENE-Fraenkel dumbbell viscosity with bead-rod dumbbell viscosity at a range of shear rate values showing deviations at small $h^*$ and $\delta Q^*_R$. Coloured lines are rodlike results at 3 different values of $h^*_\text{R}$. Symbol colour represents value of the HI parameter $h^*_\text{R}$, while symbol shape denotes different values of the extensibility $\delta Q^*_\text{R}$. Error bars are smaller than symbol size.}
  \label{hstar_changes_extensibility_scaling}
%\vskip-20pt 
\end{figure}
The molecular chemistry determines the direction of the transition dipole moment at a particular wavelength and can often be determined a-priori \cite{Norden1978}. 
Shear flow in a Couette cell is commonly used as an orientation method for LD, since it is relatively easy to measure the absorbance of a sample in a cell and only small volumes of sample are needed \cite{Rodger2009, McLachlan2013}.

In general, the overall LD signal of an oriented sample can be separated into two contributions, one from the angle the transition dipole moment makes with the molecular axis ($\alpha$) and one from the average orientation of the molecules ($S$):
\begin{equation}
    \frac{LD}{A_\text{iso}} = {LD}_\text{r} = \frac{3}{2} S (3 \cos^2{\alpha} -1)
\end{equation}
where $A_\text{iso}$ is the isotropic absorbance of the sample (prior to orientation). 
Therefore, by using experimental shear flow LD data for dilute solutions of rodlike or semi-flexible macromolecules for which $\alpha$ is known, the $S$-parameter can be extracted and compared against $S$ predictions from polymer models.

In order to reduce error bars on measured quantities, variance reduction has been used in some of the simulation results reported here.
This involves simulating a second ensemble of dumbbells at equilibrium, each of which is matched with the non-equilibrium simulation by using the same random numbers for each pair \cite{Ottinger1996}. 
When material functions are measured for the ensemble in flow, values from the equilibrium ensemble are individually subtracted from their matched dumbbell in flow, which serves to eliminate noise while keeping the same average value.

\section{Results and Discussion}

\begin{figure}[!h]
  \centerline{
  \begin{tabular}{c}
        \includegraphics[width=8.5cm,height=!]{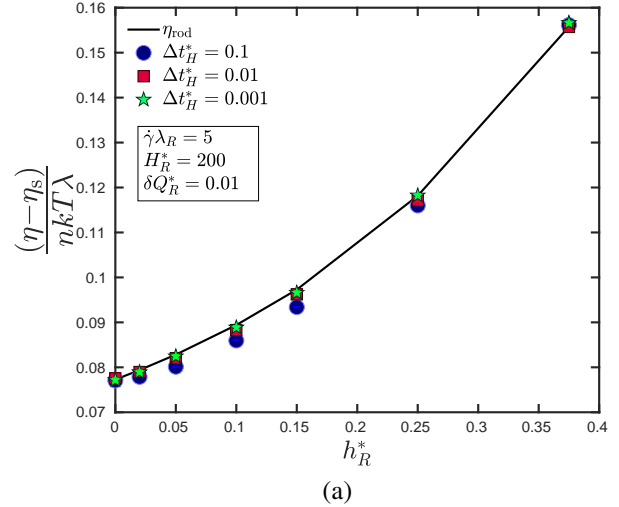} \\
        (a) \\
       \includegraphics[width=8.5cm,height=!]{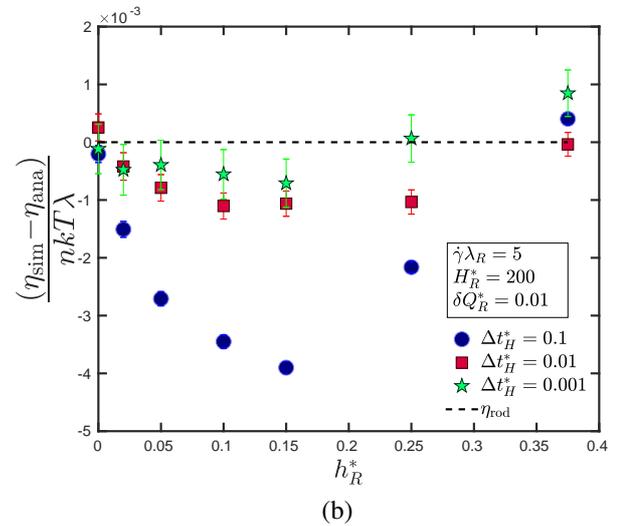} \\
       (b) \\
  \end{tabular}
  }
  \caption[Timestep convergence for low $\delta Q^*_R$]{Timestep convergence as $h^*_R$ is varied at low $\delta Q^*_R$. Plot (a) shows simulated FENE-Fraenkel viscosity for three different $\Delta t$ compared with bead-rod viscosity at the same shear rate. Plot (b) shows difference between FENE-Fraenkel and bead-rod viscosity at the same timestep values, and with the same parameters. Where not displayed, error bars are smaller than symbol size.}
  \label{timestep convergence hstar}
%\vskip-15pt  
\end{figure}

\subsection{Code Validation}

The semi-analytical solution for the rodlike distribution function was computed using MATLAB\textsuperscript{\textsuperscript{\textregistered}} ODE solvers. The results were compared with those of Stewart and Sorensen \cite{Stewart:1972gt} as well as McLachlan \textit{et al.} \cite{McLachlan2013} and found to be identical.

\begin{figure*}[t]
  \centerline{
  \begin{tabular}{c c}
        \includegraphics[width=82mm,height=!]{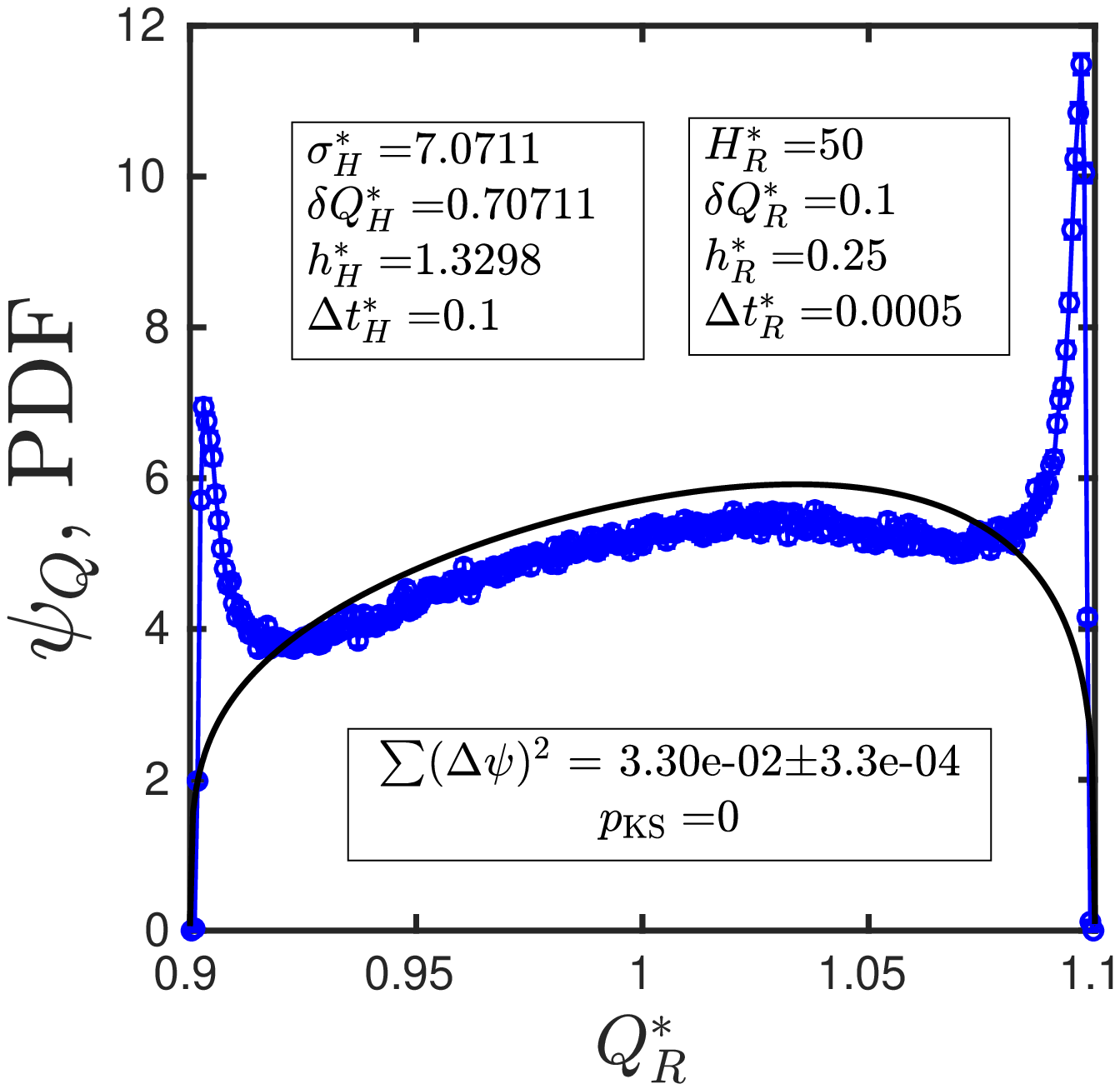} 
        & \includegraphics[width=82mm,height=!]{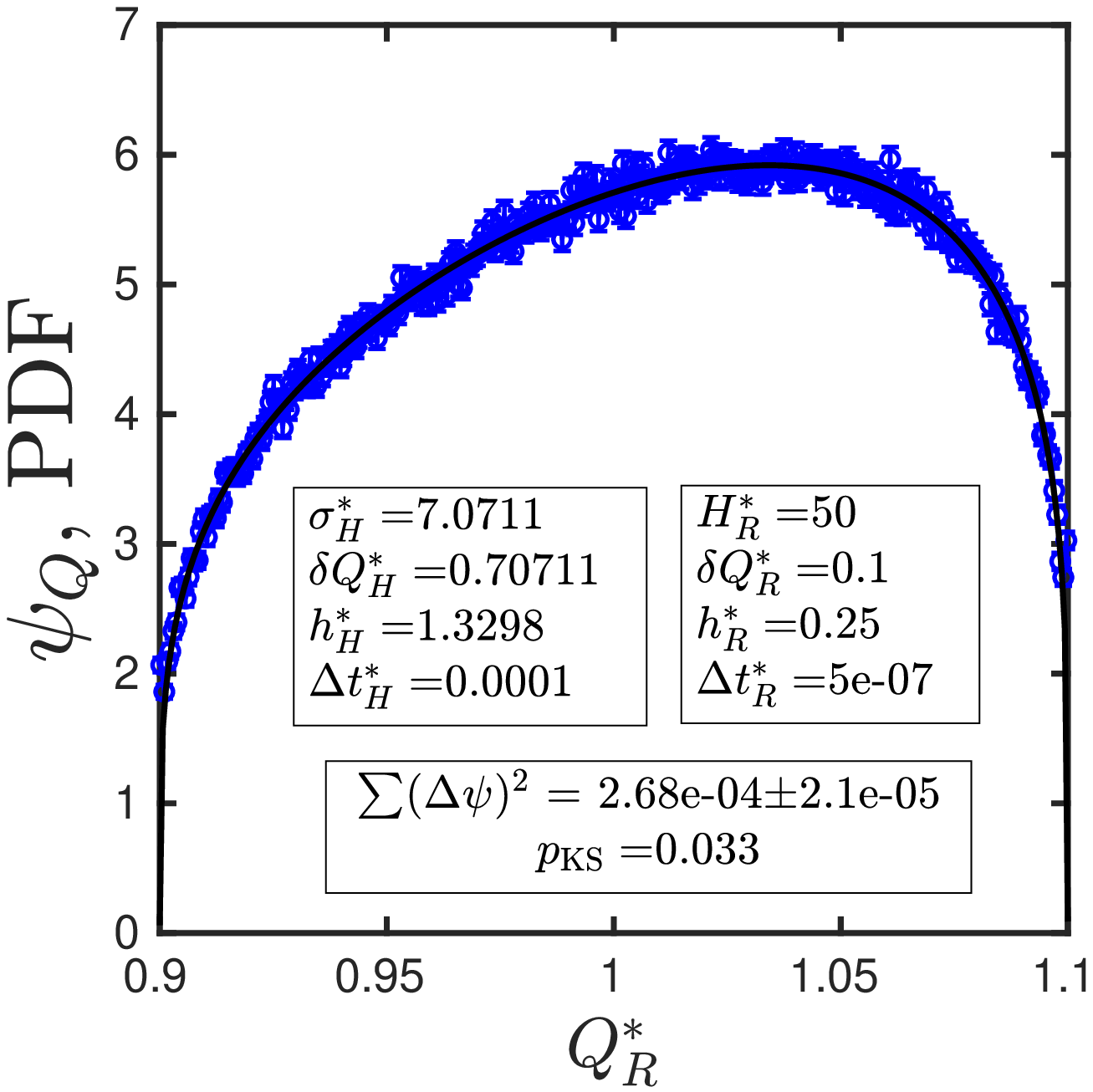} \\
        (a) & (b) \\
        \includegraphics[width=82mm,height=!]{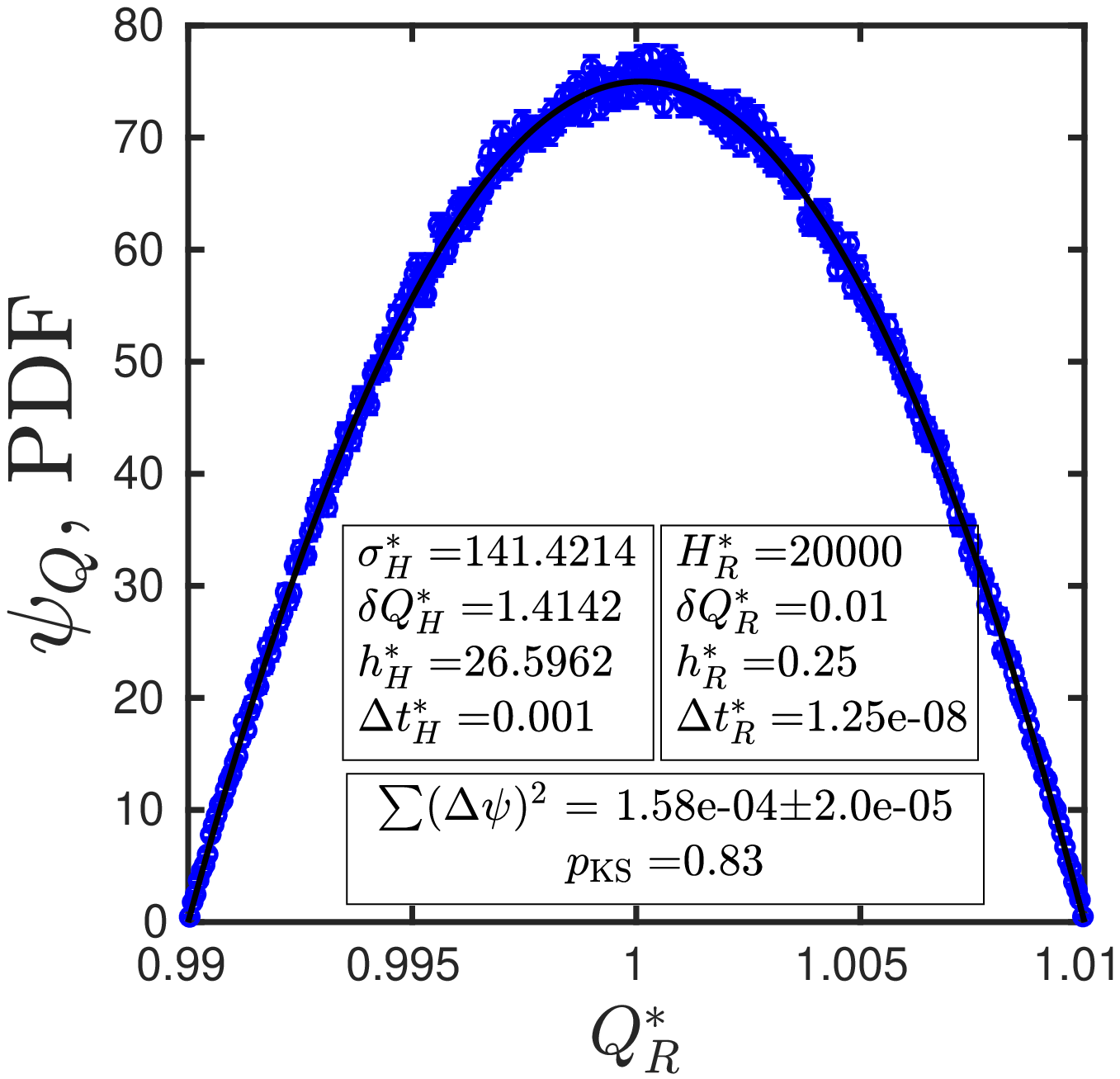} 
        & \includegraphics[width=82mm,height=!]{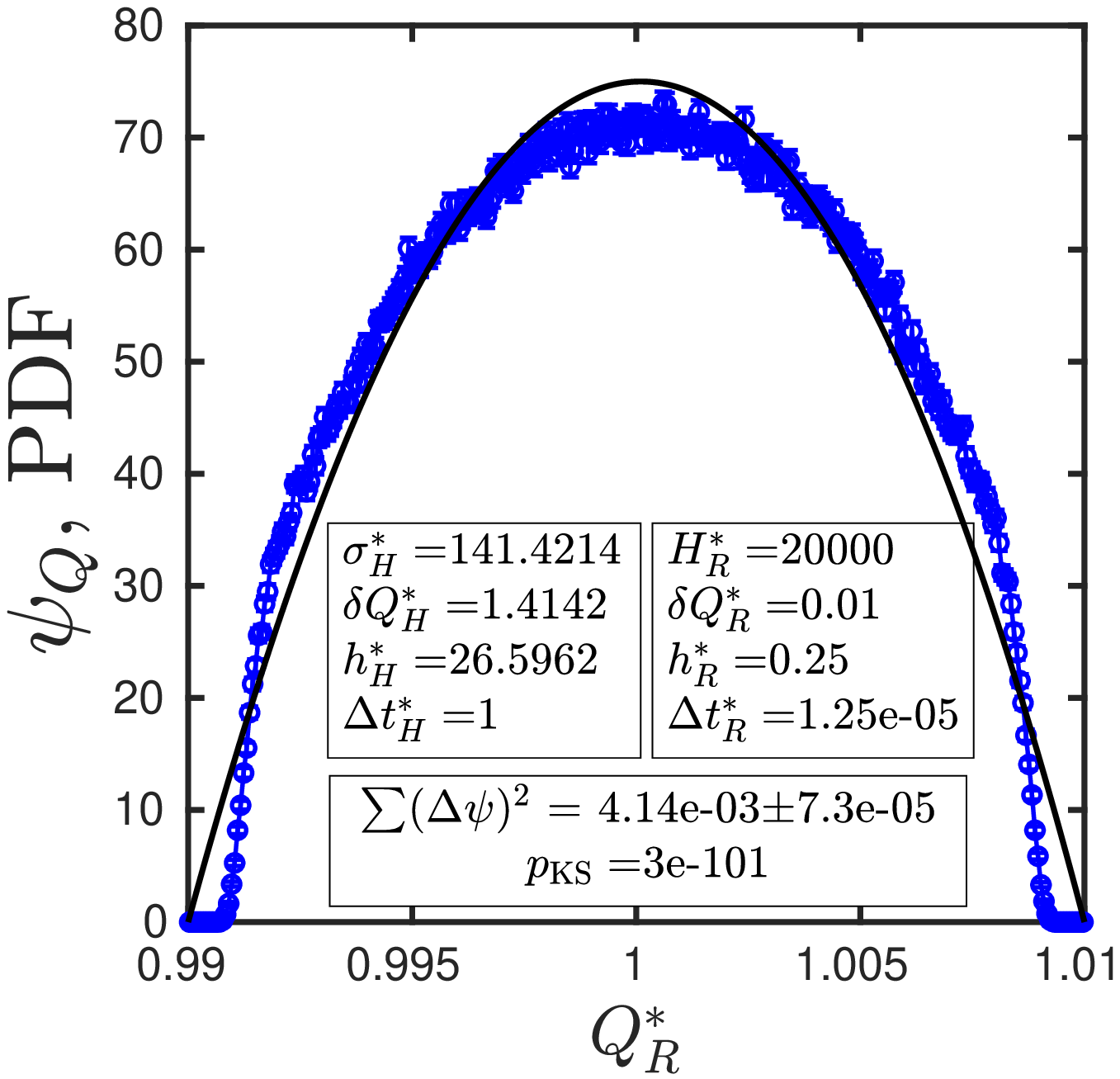} \\
        (c) & (d) \\
  \end{tabular}
  }
  \caption[Distribution functions]{Examples of distribution functions at several timestep values. The black line represents the analytical probability distribution function, while the blue circles \tikzcircle{blue} are binned dumbbell length frequencies from BD simulations. Error bars assume a Poisson distribution for each bin. Spring parameters are displayed in both rodlike and Hookean units. Summed squared error between the analytical and simulated distribution functions ($\sum (\Delta \psi)^2$, see Eq.~(31) in S.I. for expression) is displayed for each plot. The $p_\mathrm{KS}$ value is the $p$-value for a Kolmogorov-Smirnov hypothesis test, with the null hypothesis that the two distributions are the same (high $p_{KS}$ implies a closer match).}
  \label{hstar_scaling_dist_function_dQ0.1}
%\vskip-5pt  
\end{figure*}

Since there is no previous work on FENE-Fraenkel dumbbells against which to compare results, the code was validated against the expected equilibrium distribution, as well as viscometric functions of FENE dumbbells from Kailasham \textit{et al.} \cite{kailasham2018rheological}, who used the same semi-implicit predictor corrector method with a cubic polynomial solver and RPY HI.
The results are identical to within error.
Further details of these comparisons and graphs of results are given in the supporting information section 3. 

\begin{figure}[!ht]
  \centerline{
  \begin{tabular}{c}
        \includegraphics[width=8.5cm,height=!]{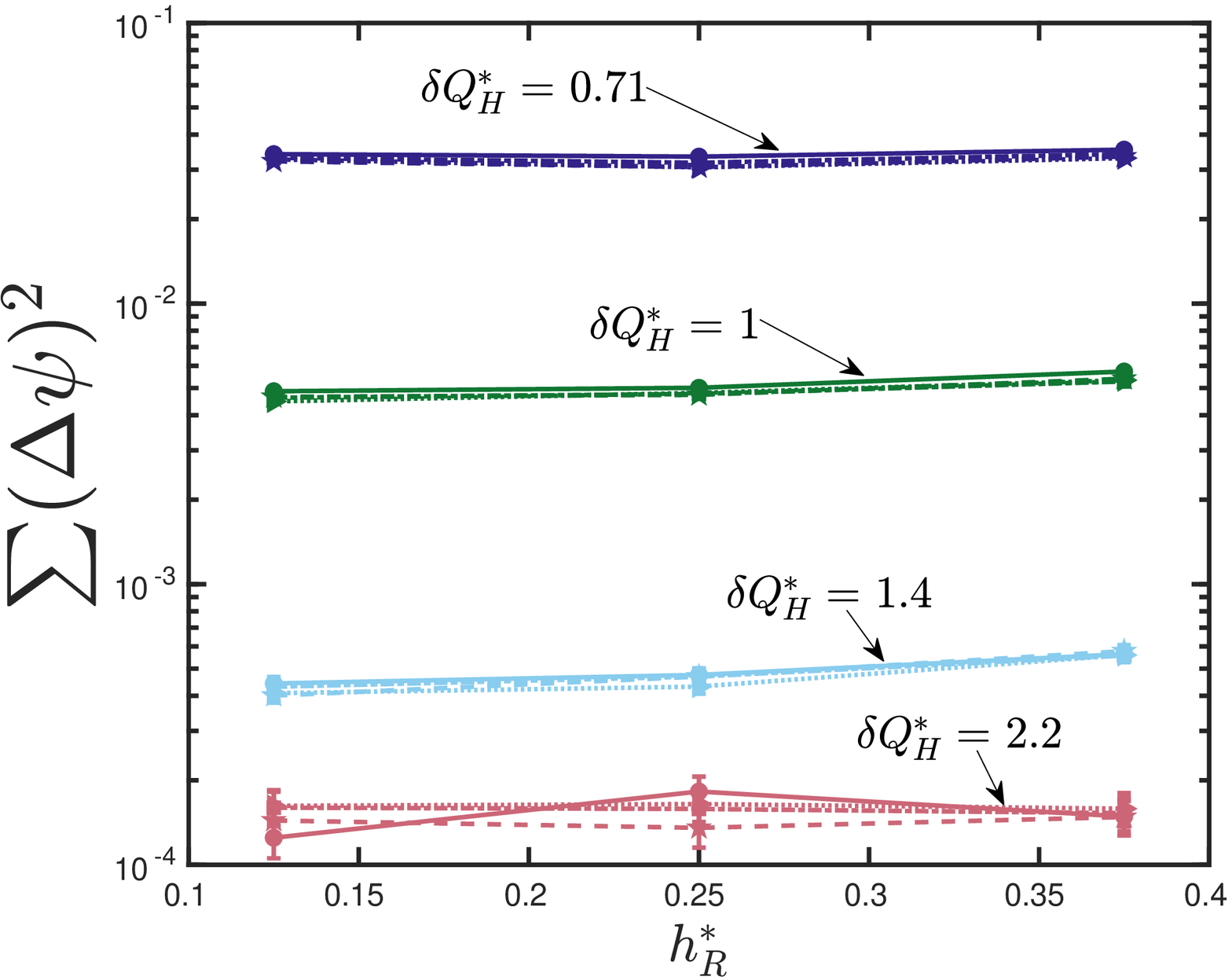} \\
        (a) \\
       \includegraphics[width=8.5cm,height=!]{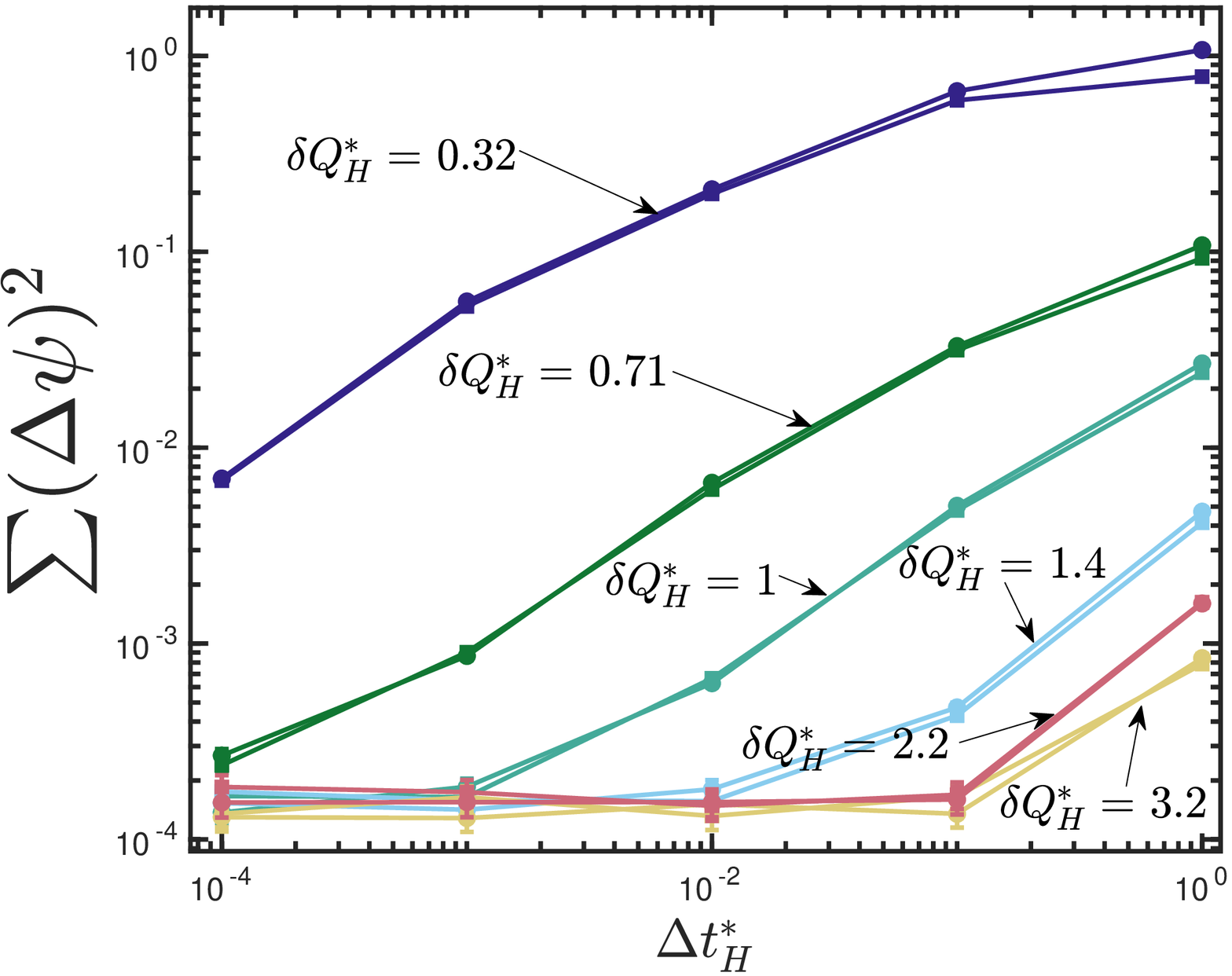} \\
       (b) \\
  \end{tabular}
  }
  \caption[Summed difference plots]{Changes in summed differences between simulated and analytical distribution functions (as seen in Fig.~\ref{hstar_scaling_dist_function_dQ0.1}) with FENE-Fraenkel spring simulation parameters. In both plots colour represents $\delta Q^*_H$ value, as labelled. (a) $\Delta t^*_H$ is held fixed at $0.1$, with full lines for $\delta Q^*_R = 0.1$, dot-dashed lines for $\delta Q^*_R = 0.05$, dashed lines for $\delta Q^*_R = 0.02$, and dotted lines for $\delta Q^*_R = 0.01$. (b) $h^*_R$ is held fixed at $0.25$, with circle symbols for $\delta Q^*_R = 0.1$ and square symbols for $\delta Q^*_R = 0.01$.  Where not displayed, error bars are smaller than symbol size.}
  \label{Difference scaling plots}
\vskip-10pt  
\end{figure}

However, there is a subtle timestep convergence issue which appears at only certain values of the hydrodynamic interaction parameter $h^*$. As seen in Fig.~\ref{hstar_changes_extensibility_scaling}, the low-shear viscosity appears nearly identical (and equal to the bead-rod value) at all values of the extensibility $\delta Q^*_R$ for $h^*_R = 0.375$, but for $h^*_R = 0.125$, the zero-shear viscosity of the FENE-Fraenkel spring appears to deviate further from the bead-rod result with decreasing extensibility, which is a counter-intuitive result.

\begin{figure}[!ht]
  \centering
  \includegraphics[width=8.5cm,height=!]{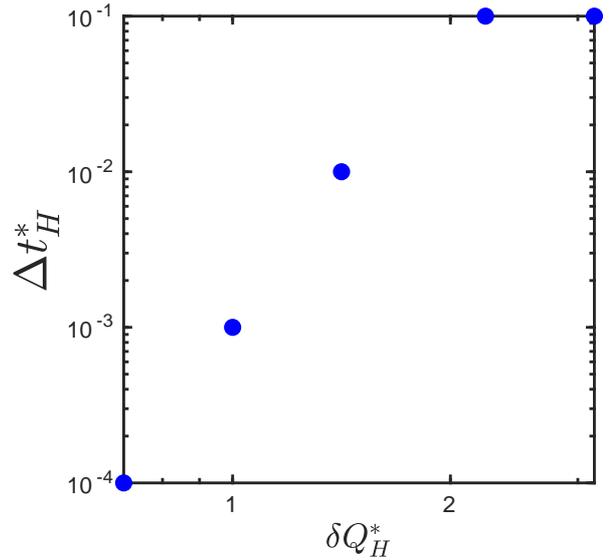}
  \caption[Timestep convergence of distribution functions]{Effect of $\delta Q^*_H$ on the $\Delta t^*_H$ required for convergence of the BD distribution function to the analytical result at equilibrium based on a Kolmogorov-Smirnov test. This is based only on the five $\Delta t$ values used in Fig.~\ref{Difference scaling plots}~(b), such that the value plotted in this figure is the highest $\Delta t^*_H$ in the set \{$10^{-4}$, $10^{-3}$, \dots, 1\} for which $p_{\mathrm{KS}} < 0.01$.  For example, a FENE-Fraenkel spring with $\delta Q^*_H = 1$ should have a timestep $\Delta t^*_H$ no larger than $10^{-3}$ for distribution function convergence.}
  \label{equilibrium timestep convergence with dQ}
\vskip-20pt    
\end{figure}

All simulations used to generate Fig.~\ref{hstar_changes_extensibility_scaling} used the same $\Delta t^*_H = 0.2$ (equivalently $\Delta t^*_R = 0.00025$), which was small enough to ensure timestep convergence for $h^*_R = 0.375$. However, as can be seen in Fig.~\ref{timestep convergence hstar}, timestep convergence actually varies considerably and non-monotonically with the value of $h^*$. 
Specifically, at $h^*_R = 0$ and $h^*_R = 0.375$ (representing no HI or osculating beads respectively), a change from $\Delta t^*_H = 0.1$ to $\Delta t^*_H = 0.001$ makes very little difference to the measured viscosity, while for $h^*_R = 0.15$, the viscosity varies significantly.
While one might expect that this is due to some numerical error at intermediate $h^*$, comparison of the simulated probability distribution function of dumbbell lengths $\psi(Q)$ at equilibrium with the analytical result seems to suggest that the convergence at low and high $h^*$ is a simple coincidence, as will be shown below.

Fig.~\ref{hstar_scaling_dist_function_dQ0.1} gives some examples of these distribution function plots. The sum of squared differences between the simulated and analytical distributions is denoted by $\sum (\Delta \psi)^2$ (the precise definition of which is given in Eq. 31 of the supporting information), while the $p_\mathrm{KS}$ value reported in the figure is the $p$-value for a Kolmogorov-Smirnov hypothesis test, with the null hypothesis that the two distributions are the same. High $p_{KS}$ implies a closer match (details are given in the supporting information section 3.2). Figs.~\ref{hstar_scaling_dist_function_dQ0.1} (a) and (c) are not converged, with a $p_\mathrm{KS} < 0.01$, while Figs.~\ref{hstar_scaling_dist_function_dQ0.1} (b) and (d) are converged (based on $p_\mathrm{KS} > 0.01$). By examining these distribution function plots for a range of possible FENE-Fraenkel spring parameters and timestep widths, it appears that convergence depends primarily upon the balance between $\delta Q^*_H$ and $\Delta t^*_H$, with other parameters playing a minor role 
(Note that in non-dimensional form, the FENE-Fraenkel spring is characterised by two parameters, as can be seen by manipulating Eq.~\eqref{FF_force_eqn} with Eq.~\eqref{QH to QR conv}, and so $\delta Q^*_R$ and $\delta Q^*_H$ are sufficient to fully characterise the spring force). This can be seen clearly in Fig.~\ref{Difference scaling plots}, which plots the summed difference for a wide range of FENE-Fraenkel springs and timestep widths. From Fig.~\ref{Difference scaling plots} (a) with constant $\Delta t^*_H$ and varied $h^*_R$, $\delta Q^*_H$ and $\delta Q^*_R$, it's clear that for a particular timestep, the $\delta Q^*_H$ value is by far the most impactful parameter, with $h^*_R$ and $\delta Q^*_R$ having only minor effects. Fig.~\ref{Difference scaling plots} (b) then plots the timestep convergence for constant $h^*_R$, showing a levelling off of the summed difference at low $\Delta t^*_H$.

The conclusion is that we can be reasonably certain of equilibrium timestep convergence of the distribution functions if we choose a sufficiently small $\Delta t^*_H$ given a particular $\delta Q^*_H$, with other parameters being relatively unimportant.
Fig.~\ref{equilibrium timestep convergence with dQ} quantifies this relationship, giving the approximate $\Delta t^*_H$ required for $p_\mathrm{KS} > 0.01$ in the Kolmogorov-Smirnov test such that the simulated distribution is the same as the analytical distribution at a particular value of $\delta Q^*_H$.
A $\Delta t^*_H$ larger than that in the figure may give unpredictable results, as evidenced by Fig.~\ref{timestep convergence hstar}. 
Note, however, that a value of $\Delta t^*_H$ smaller than that in Fig.~\ref{equilibrium timestep convergence with dQ} is necessary but not sufficient to guarantee convergence away from equilibrium.
Particularly at very high shear rates, a smaller $\Delta t^*_H$ may be required.

Notably, this issue may be the reason Larson and coworkers \cite{Hsieh2006} were unable to reproduce the zero-shear viscosity of a bead-rod model with a FENE-Fraenkel spring when HI is included, since they used $\delta Q^*_H < 0.1$ and $\Delta t^*_H = 4$, which appears too large to ensure timestep convergence on the distribution functions. 
It would be worth revisiting the problem for the bead-spring-chain to determine if this is the case.

\subsection{Comparison of Rodlike Models with FENE-Fraenkel Dumbbell}

\begin{figure}[t]
  \centering
  \includegraphics[width=8cm,height=!]{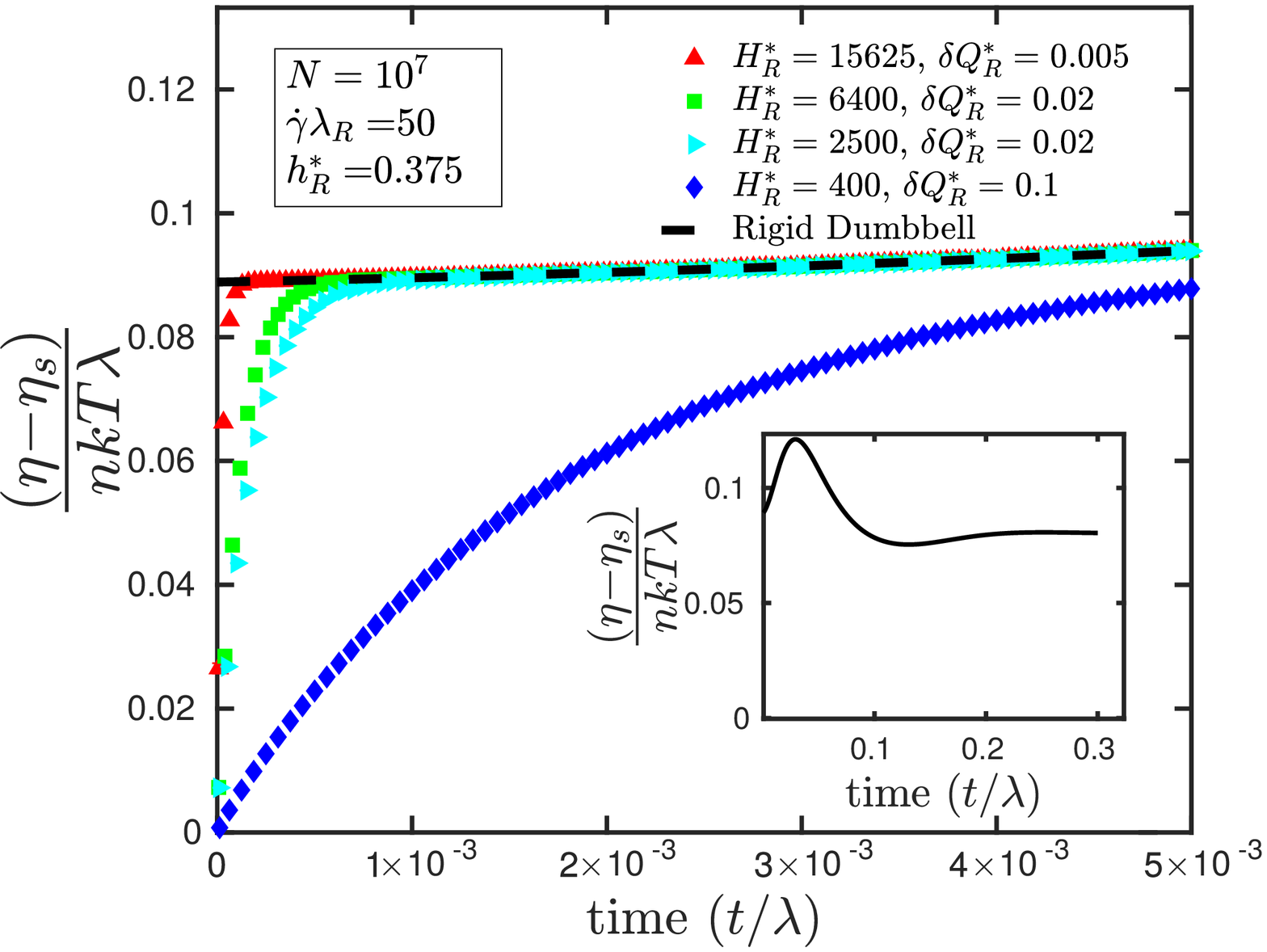}
  \caption[FENE-Fraenkel spring stress jump]{Stress jump in a FENE-Fraenkel spring at various values of spring stiffness and extensibility. Inset shows full transient viscosity curve for bead-rod dumbbell, which FENE-Fraenkel simulations follow accurately after initial stress jump. Error bars are smaller than symbol size.}
  \label{Stress Jump}
%\vskip-10pt    
\end{figure}

One of the key aims of using the FENE-Fraenkel spring is to reproduce the rheological behaviour of a rigid rod. 
In this section the material properties of a rigid dumbbell are compared to those of a FENE-Fraenkel-spring dumbbell simulated using BD. 
In general, the rodlike system of non-dimensionalisation is adopted for the FENE-Fraenkel dumbbell results, which enables direct comparisons between bead-spring and bead-rod results without normalisation or re-scaling of variables. 
Note that the spring stiffness and extensibility $\delta Q^*_R $ and $H^*_R$ are chosen such that $\delta Q^*_H = \sqrt{2}$ or $\sqrt{5}$ (i.e. $\delta Q^*_H > 1$), so that when $\Delta t^*_H < 10^{-2}$ the underlying length distribution function is timestep converged.
These choices of $\delta Q^*_H$ give a reasonable balance between required computational time and accuracy with respect to reproducing bead-rod results.
The effects of varying $\delta Q$, $H$ and $\sigma$ systematically will be investigated in later sections.

In the following discussion, the term `accuracy' is used to refer to how well the FENE-Fraenkel spring reproduces the bead-rod results.
For example, a lower spring extensibility is said to give a more accurate viscosity than a spring with higher extensibility if the simulated viscosity is closer to the bead-rod viscosity for the lower extensibility.

\subsubsection{Stress Jump}

Comparing the expression for the stress tensor of bead-rod dumbbells in Eq.~\eqref{Bead-Rod-Dumbbell_stress_tensor_with_HI} with that of bead-spring dumbbells in Eq.~\eqref{Bead-Spring-Dumbbell_stress_tensor}, both expressions share a term which varies with $\langle \bm{u} \bm{u} \rangle$, since $\bm{Q}$ and $\bm{F}$ are both directed along $\bm{u}$. 
If flow is switched on at $t=0$, this term must be isotropic (i.e. equal to the unit tensor $\bm{\delta}$) at $t \leq 0$, since the dumbbells will be in their equilibrium configuration prior to flow and at the first instant.
Therefore, the off-diagonal elements of the stress tensor must be uniformly 0 at the inception of flow for any form of the spring potential.
On the other hand, the stress tensor for the rod contains an additional term, $\{\bm{\kappa} \bm{:} \langle \bm{u u u u} \rangle \}$, often referred to as the `viscous' contribution to the stress tensor \cite{doyle1997dynamic}, which scales with the flow tensor and the fourth power of the dumbbell orientation.
As this term has non-zero cross-terms (i.e. $\{\bm{\kappa} \bm{:} \langle \bm{u u u u} \rangle \}_{x,y} \neq 0$) for an equilibrium $\psi(\bm{u})$, the viscosity of a rod is non-zero at $t=0$, giving an instantaneous `stress jump' at the inception of shear flow.
This seems to be a fundamental difference between unconstrained bead-spring and constrained bead-rod models.

\begin{figure}[t]
  \centering
  \includegraphics[width=8.5cm,height=!]{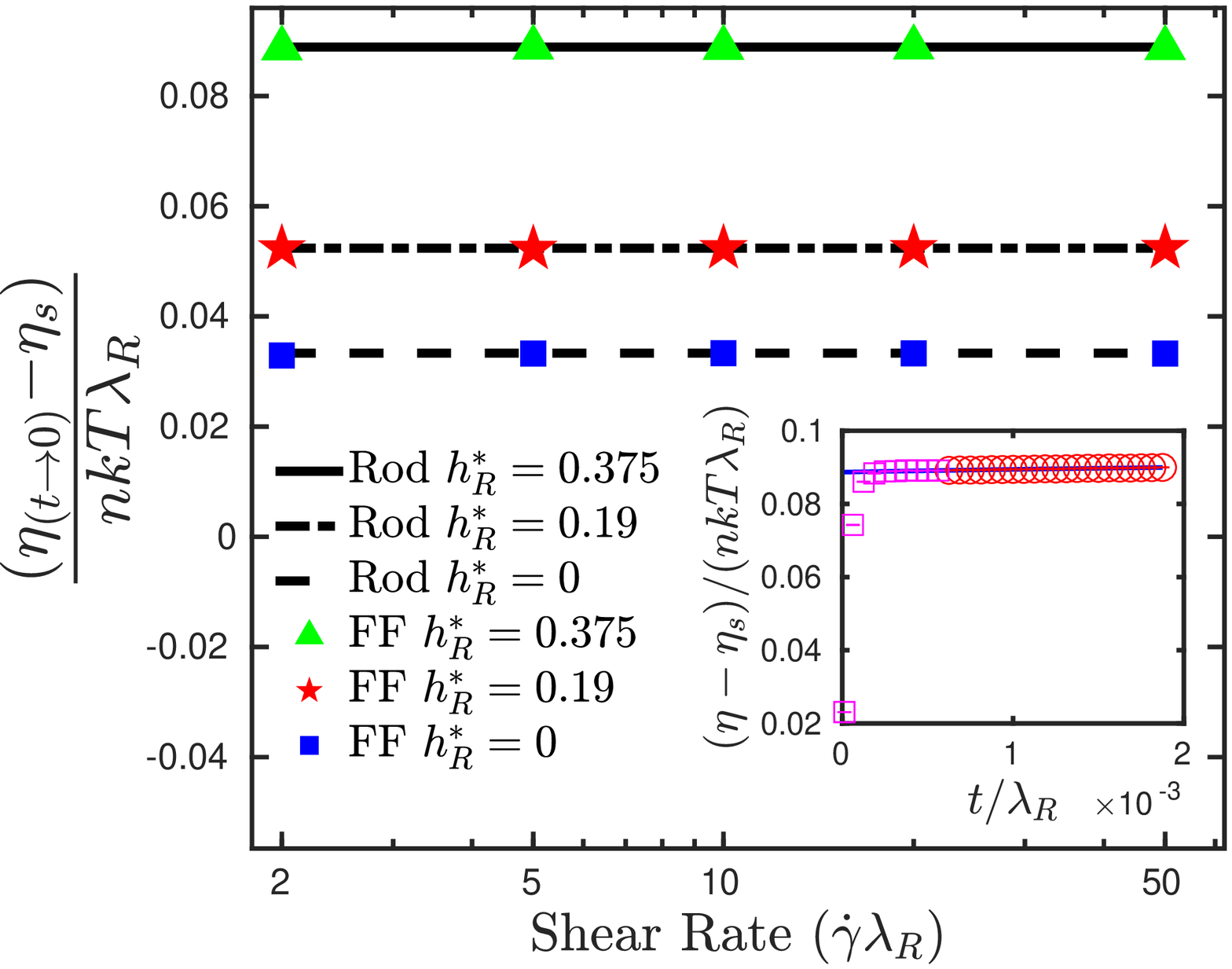}
  \caption[Stress jump is independent of shear rate]{Shear rate independence of stress jump in both bead-rod dumbbells and FENE-Fraenkel spring dumbbells. FENE-Fraenkel springs have $H^*_R = 400$ and $\delta Q^*_R = 0.005$. Values for the FENE-Fraenkel dumbbell were calculated using a fourth-order polynomial fit extrapolated to zero time, as shown in the inset. The rigid dumbbell values are given directly by Eq.~\eqref{Bead-Rod-Dumbbell_stress_tensor_with_HI}. In the inset, Red circles \tikzcircle{red} represent data points used for the extrapolation, while magenta squares \tikzsquare{magenta} represent data obtained from BD simulations but not used in the extrapolation, since the viscosity had not leveled off to the linear region. Blue line is a 4th-order polynomial fit to the red circles. Error bars are smaller than symbol size.}
  \label{Stress jump doesn't vary with sr}
\vskip-10pt  
 \end{figure}

In spite of the absence of this term for FENE-Fraenkel springs, they were found to exhibit a `pseudo-stress-jump', with an extremely rapid rise in viscosity to match that of bead-rod dumbbells at the inception of flow. Fig.~\ref{Stress Jump} shows BD simulation results for the transient behaviour of FENE-Fraenkel springs as flow is switched on. The sets of parameters used in Fig.~\ref{Stress Jump} all lead to an asymptotic convergence to the bead-rod results on a timescale which is a small fraction of the total time to reach steady-state (steady state is reached at $t \approx 0.25 \lambda_R$ for $\dot{\gamma} \lambda R = 50$, as displayed in Fig.~\ref{Stress Jump} inset).
As one would intuitively expect, this pseudo-stress-jump occurs over a shorter period of time when the spring stiffness is increased or the extensibility is decreased.

For bead-rod dumbbells, the magnitude of the stress jump should be independent of shear rate but still vary with the hydrodynamic interaction parameter, as predicted by theory \cite{bird1987dynamics}. 
Fig.~\ref{Stress jump doesn't vary with sr} shows that this relation holds for FENE-Fraenkel dumbbells when the viscosity is extrapolated to $t=0$. An example of this extrapolation is shown in the inset of Fig.~\ref{Stress jump doesn't vary with sr}, where the data points used for a polynomial fit are chosen in the approximately linear region immediately after the initial rapid stress jump.

Note that in the context of experimental measurements, the fact that this stress jump is not strictly instantaneous is mostly irrelevant, since it is generally on the order of microseconds and can be made arbitrarily small by decreasing the extensibility or increasing the spring stiffness. 
For example, later in section \ref{experimental comparisons LD} it will be shown that an $800$ nm contour length bacteriophage with an aspect ratio of $\approx 100$ can be reasonably modelled by a FENE-Fraenkel spring with $a = 20.9$ nm, $\sigma = 727$ nm, $H^*_R = 200$, $\delta Q^*_R = 0.1$ and hence $\delta Q \approx 73$. For this parameter set, the stress jump occurs over approximately $300$ $\upmu$s, which is shorter than the step-change time of most widely used rheometers (generally on the order of milliseconds). Therefore, although the FENE-Fraenkel spring is not a perfect reproduction of the rodlike stress jump, it should be able to reproduce any experimental measurement of the stress jump given appropriately chosen parameters.

\begin{figure}[t]
  \centerline{
  \begin{tabular}{c}
        \includegraphics[width=8.5cm,height=!]{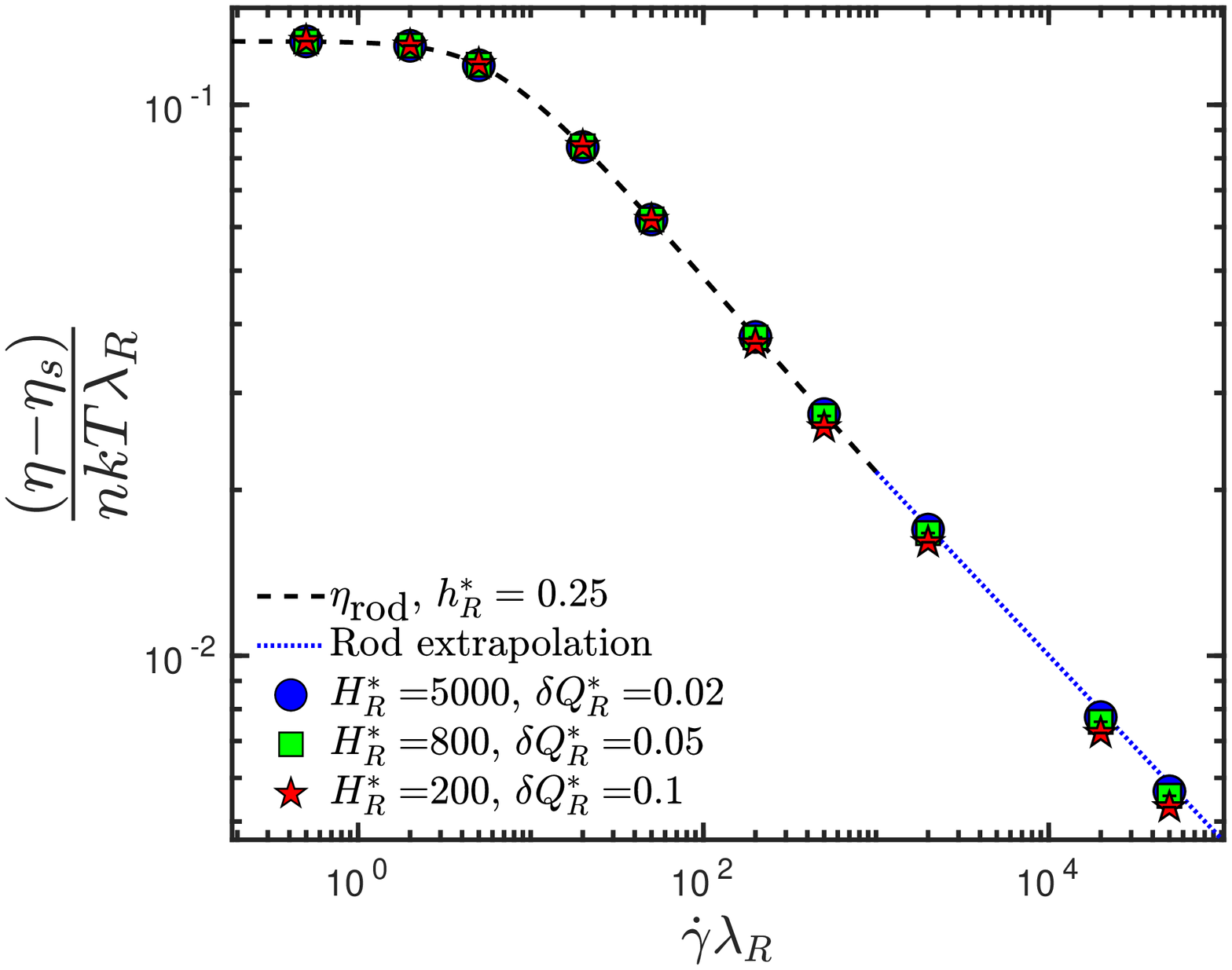} \\
        (a) \\
       \includegraphics[width=8.5cm,height=!]{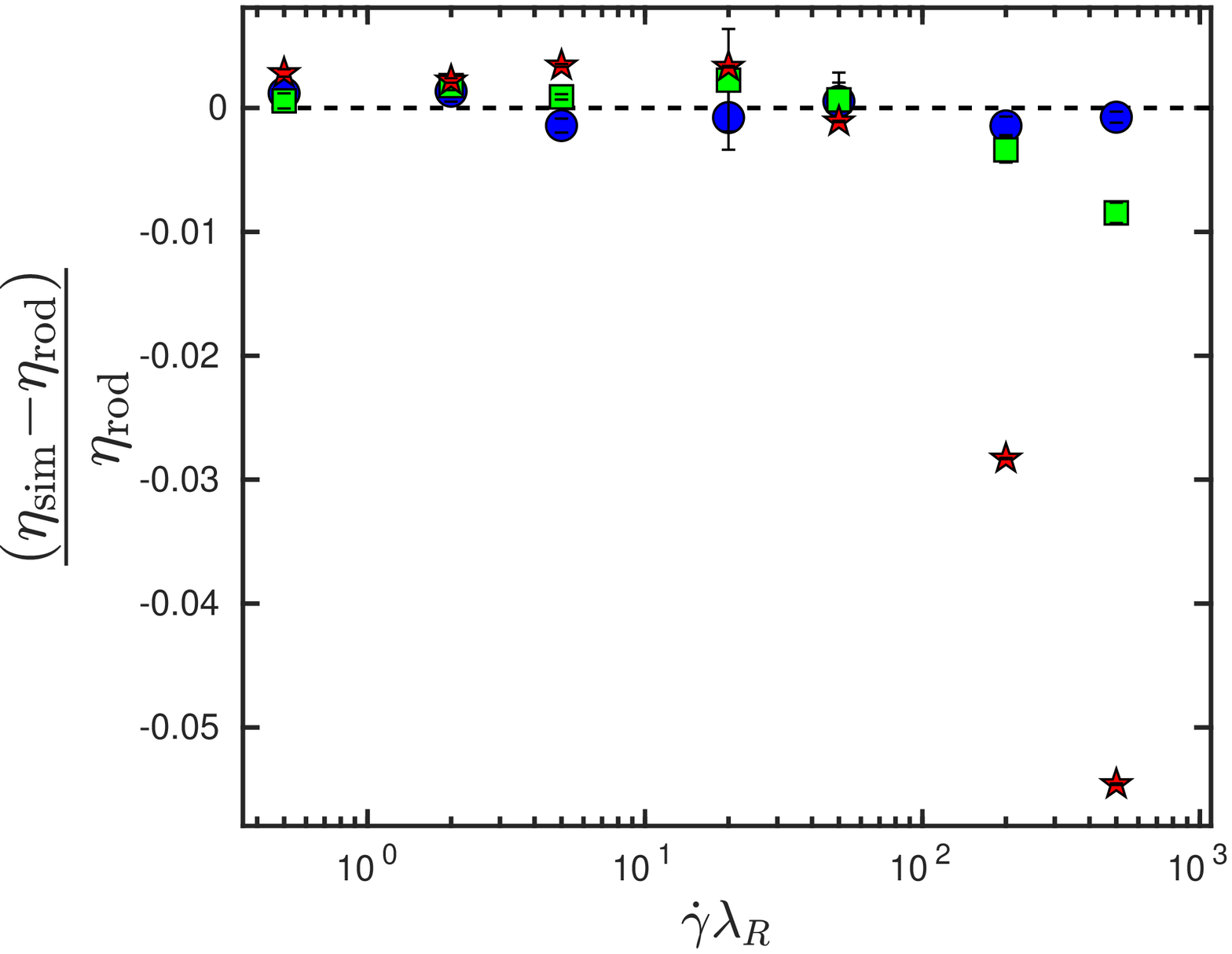} \\
       (b) \\
  \end{tabular}
  }
  \caption[Viscosity comparison FF spring and Rod]{Non-dimensional viscosity against non-dimensional shear rate for bead-rod dumbbells (dashed lines) and bead-FENE-Fraenkel-spring dumbbells (symbols). Error bars are smaller than symbol size. Fig.~(a) shows direct comparison, while Fig.~(b) gives the relative error between the BD simulations and rodlike semi-analytical result, with the symbols having the same meaning in both (a) and (b). Blue dotted line is a power-law extrapolation from the tail of the bead-rod curve, with exponent $-1/3$.}
  \label{FF_Rigid_comparison_eta}
\vskip-10pt  
\end{figure}

\begin{figure}[!ht]
  \centerline{
  \begin{tabular}{c}
        \includegraphics[width=8.5cm,height=!]{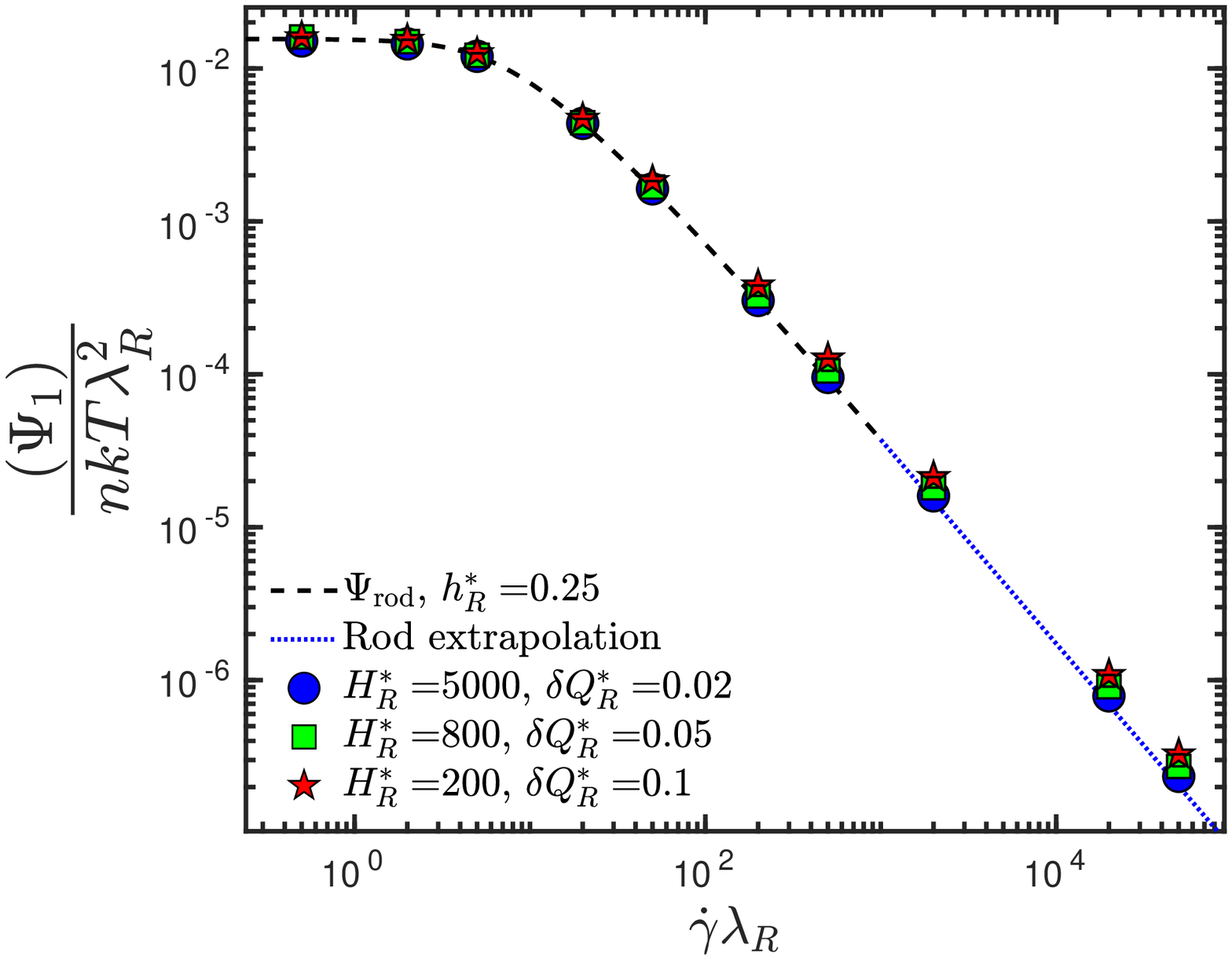} \\
        (a) \\
       \includegraphics[width=8.5cm,height=!]{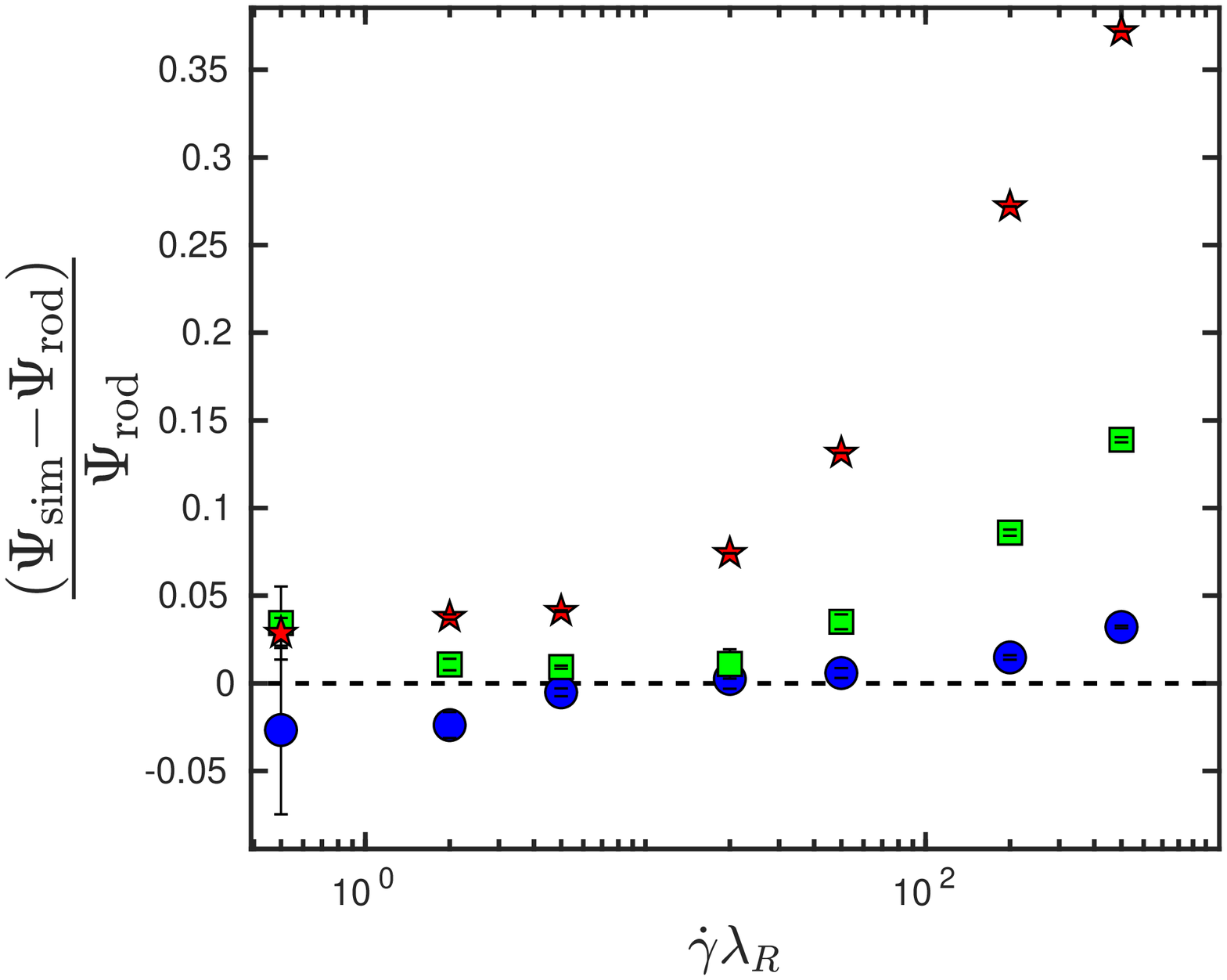} \\
       (b) \\
  \end{tabular}
  }
  \caption[Viscosity comparison FF spring and Rod]{Non-dimensional first normal stress difference against non-dimensional shear rate for bead-rod dumbbells (dashed lines) and bead-FENE-Fraenkel-spring dumbbells (symbols). Error bars are smaller than symbol size. Fig.~(a) shows direct comparison, while Fig.~(b) gives the relative error between the BD simulations and rodlike semi-analytical result, with the symbols having the same meaning in both (a) and (b). Blue dotted line is a power-law extrapolation from the tail of the bead-rod curve, with exponent $-4/3$.}
  \label{FF_Rigid_comparison_Psi1}
\vskip-10pt 
  \end{figure}

\begin{figure}[!ht]
  \centerline{
  \begin{tabular}{c}
        \includegraphics[width=8.5cm,height=!]{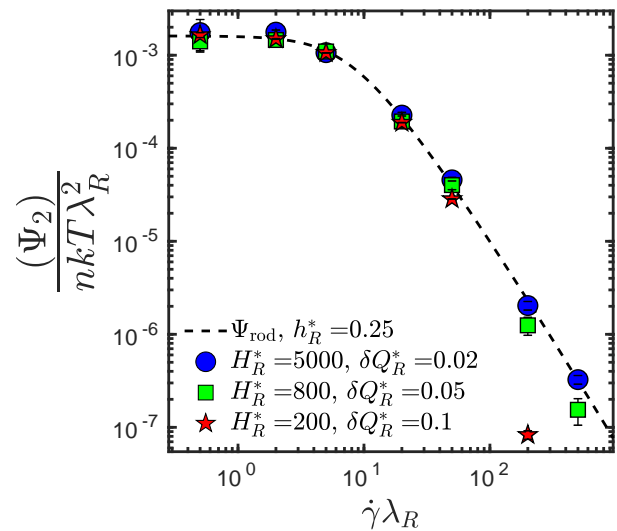} \\
        (a) \\
       \includegraphics[width=8.5cm,height=!]{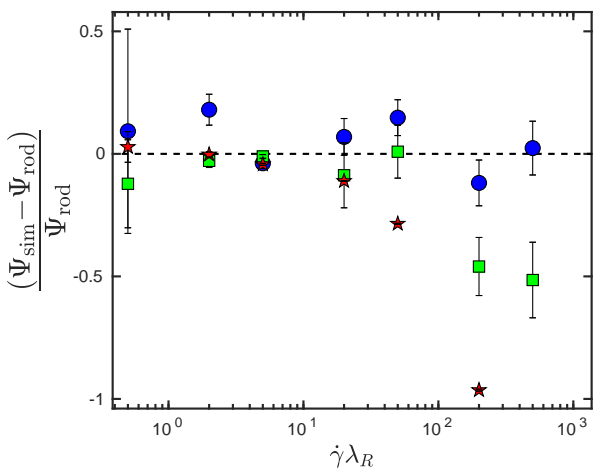} \\
       (b) \\
  \end{tabular}
  }
  \caption[Viscosity comparison FF spring and Rod]{Non-dimensional second normal stress difference against non-dimensional shear rate for bead-rod dumbbells (dashed lines) and bead-FENE-Fraenkel-spring dumbbells (symbols). Error bars are smaller than symbol size. Fig.~(a) shows direct comparison, while Fig.~(b) gives the relative error between the BD simulations and rodlike semi-analytical result, with the symbols having the same meaning in both (a) and (b). $\Psi_2$ is not displayed beyond $\dot{\gamma} \lambda_R = 500$ as error is too large to give meaningful results.}
  \label{FF_Rigid_comparison_Psi2}
\vskip-10pt 
\end{figure}

\subsubsection{Material Functions} \label{material function results}
Figs. \ref{FF_Rigid_comparison_eta}~(a) to \ref{FF_Rigid_comparison_Psi2}~(a) show the scaling of material functions $\eta$, $\Psi_1$ and $\Psi_2$ with shear rate for the FENE-Fraenkel spring at various values of the spring stiffness $H^*_R$ and extensibility $\delta Q^*_R$.
Note that the semi-analytical solution method for bead-rod dumbbells is unstable beyond $\dot{\gamma} \lambda_R = 10^3$ (since an increasingly ill-conditioned matrix at high shear rates must be inverted), and so bead-rod curves are only displayed up to this shear rate.
Power law extrapolations are provided as a guide to the eye for higher shear rates.
Figs. \ref{FF_Rigid_comparison_eta}~(b) to \ref{FF_Rigid_comparison_Psi2}~(b) also show the relative error between the bead-rod dumbbell material functions and the FENE-Fraenkel spring, as it is hard to discern the difference on the log scales.
Although results are given only for $h^*_R = 0.25$, plots for other values of the HI parameter give qualitatively similar results with respect to the scaling of viscometric functions as $H^*_R$ and $\delta Q^*_R$ are varied. 

\begin{figure}[tbh]
  \centerline{
  \begin{tabular}{c}
        \includegraphics[width=8.5cm,height=!]{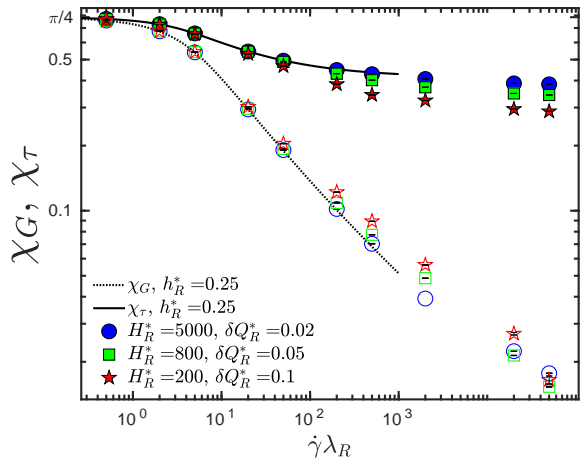} \\
        (a) \\
       \includegraphics[width=8.5cm,height=!]{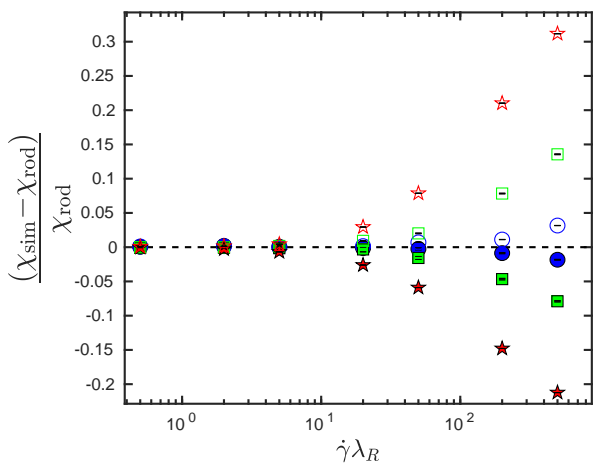} \\
       (b) \\
  \end{tabular}
  }
  \caption[$\chi_G$ and $\chi_{\tau}$ scaling with shear rate]{Black lines show $\chi_G$ (dotted line) and $\chi_{\tau}$ (filled line) for a bead-rod dumbbell. Filled in shapes are FENE-Fraenkel simulation averages for $\chi_{\tau}$, while open shapes are those for $\chi_G$. Fig.~(a) gives a direct comparison of BD simulations with rodlike semi-analytical calculations, while Fig.~(b) shows the relative error between the two at each shear rate, with the symbols having the same meaning in both (a) and (b).}
  \label{chi scaling rodlike}
  \vskip-10pt 
\end{figure}

These results show that the FENE-Fraenkel spring is able to reproduce the shear-rate-scaling behaviour of all three measured viscometric functions over three decades of shear rates given appropriately chosen spring parameters.
A stiffer spring (with larger $H^*_R$ and/or smaller $\delta Q^*_R$) appears to uniformly give more accurate results for all shear rates.
At high shear rates, the material properties of the less stiff spring ($H^*_R = 200$, $\delta Q^*_R = 0.1$) deviate significantly from the bead-rod results, while at lower shear rates there is less difference.
It appears that the `spring-like' nature of the FENE-Fraenkel dumbbells is being revealed at high shear rates, which causes a deviation from the $-1/3$ power law scaling in the viscosity. 

For the FENE-Fraenkel dumbbell, the steady-shear second normal stress difference is strictly positive for the parameter set used here, matching with the bead-rod predictions as in Fig.~\ref{FF_Rigid_comparison_Psi2}.
While this is consistent with the previous calculations of Stewart and Sorenson \cite{Stewart:1972gt}, note that this behaviour is different from that of springs.
For example, Hookean springs give a negative $\Psi_2$ when fluctuating hydrodynamic interactions are included \cite{Prabhakar2002}, and it is this Hookean behaviour which is consistent with careful experimental measurements of polymer solutions \cite{keentok1980measurement}.
It will later be shown that the FENE-Fraenkel spring does in fact give a negative $\Psi_2$ for sufficiently `spring-like' sets of parameters.

The first and second normal stress differences of the FENE-Fraenkel spring show considerably more deviation from the rodlike results than the viscosity for all spring parameters.
This is particularly noticeable at high shear rates, such that to obtain a maximum of $10 \%$ error in the measured material parameters with respect to the rodlike results requires $H^*_R > 5000$, $\delta Q^*_R < 0.02$ for $\Psi_2$ in Fig.~\ref{FF_Rigid_comparison_Psi2}, and $H^*_R > 800$, $\delta Q^*_R < 0.05$ for $\Psi_1$ in Fig.~\ref{FF_Rigid_comparison_Psi1}, while only $H^*_R > 200$, $\delta Q^*_R < 0.1$ in Fig.~\ref{FF_Rigid_comparison_eta} for $\eta$. 
There appears to be no universal `safe' set of parameters to ensure convergence to rodlike results, with the degree of accuracy instead depending on the shear rate and measured observable.

\subsubsection{$\chi_G$ and $\chi_{\tau}$ scaling with shear rate} \label{chi rods}
As previously discussed, the two $\chi$ parameters, $\chi_G$ and $\chi_{\tau}$, represent the orientation of the gyration tensor and the stress tensor respectively. 
These should be identical at equilibrium (or at sufficiently low shear rates), since when there is no shear the stress tensor is directly proportional to the gyration tensor.
In general, at higher shear rates $\chi_G$ and $\chi_\tau$ may separate, indicating that these two tensors are no longer directly proportional and hence the stress optical rule no longer holds.

Fig.~\ref{chi scaling rodlike} compares simulation results for FENE-Fraenkel dumbbells with the semi-analytical bead-rod solution. 
A lower extensibility or higher stiffness seems to uniformly give more accurate results over the whole range of shear rates for both $\chi_\tau$ and $\chi_G$. 
For the spring parameters chosen here, the stress-optical rule clearly does not hold, as for the bead-rod dumbbell. 
Furthermore, all three parameter sets show the same qualitative behaviour as a rod, with a continuously decreasing $\chi_G$ and a plateau in $\chi_\tau$ at high shear rates.
This plateau is a feature of the power-law scaling of $\eta$ and $\Psi_1$, as follows from Eq.~\eqref{chi tau definition}.
Since $\eta$ has a $\approx -1/3$ scaling, while $\Psi_1$ has a $\approx -4/3$ scaling, $2 \eta_p / \Psi_1 \dot{\gamma}$ will be approximately constant at high shear rates, and hence $\chi_\tau$ will also be constant.
This is not the general behaviour of a Hookean or FENE spring, which display a power-law decay in both $\chi_G$ and $\chi_\tau$, as will be seen in section \ref{chi with extensibility}.

\subsection{Effects of increased extensibility}
The FENE-Fraenkel spring is able to not only represent a rod in the limit of low $\delta Q$ and high $H$, but also an entropic spring such as a FENE spring in the limit of $\sigma \rightarrow 0$. 
The range of possible $\sigma$, $\delta Q$ and $H$ values between these two limits can therefore mimic a variety of possible force-extension relations.

\begin{figure}[t]
  \centering
    \begin{tabular}{c}
        \includegraphics[width=80mm]{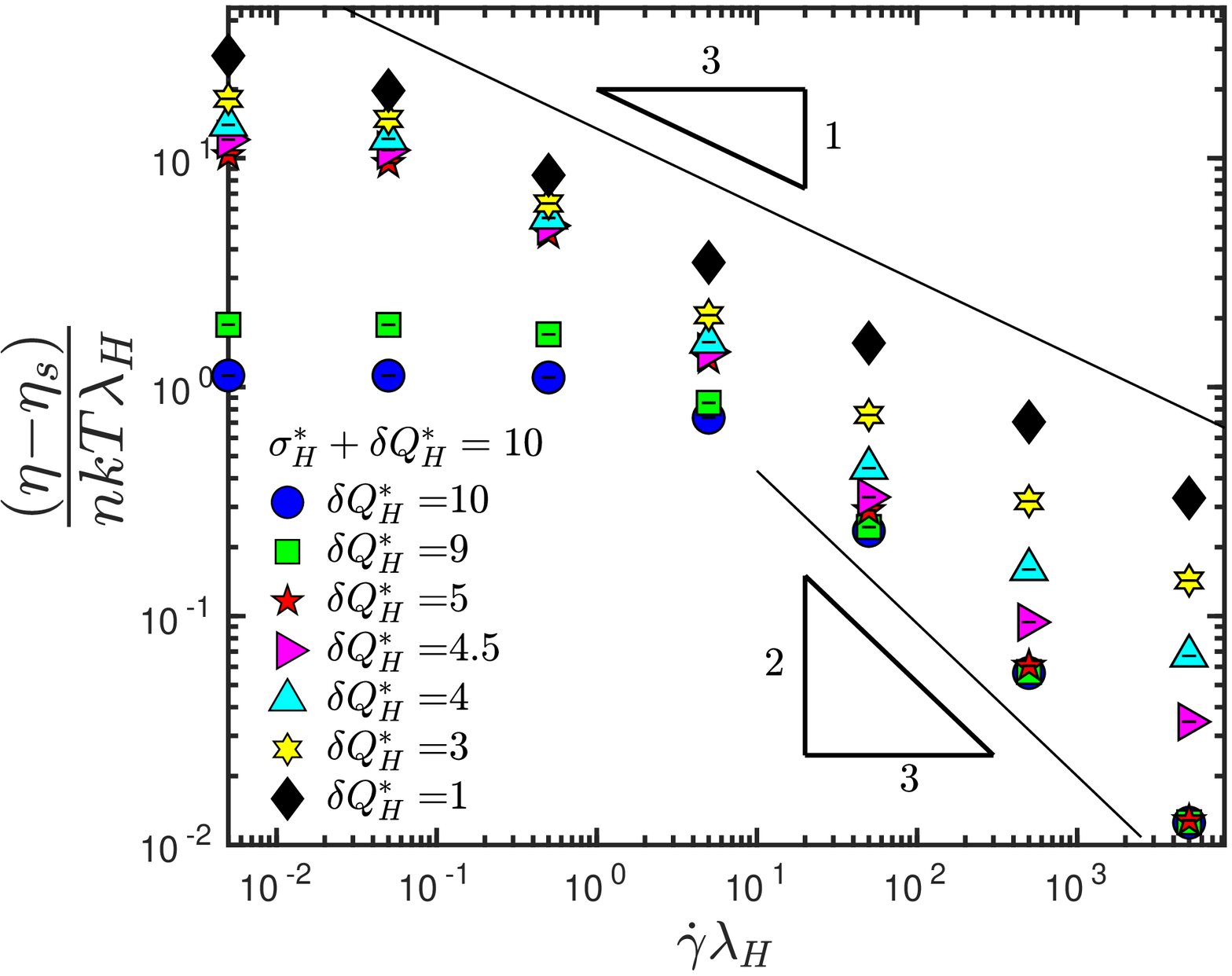} \\
        (a) \\
       \includegraphics[width=80mm]{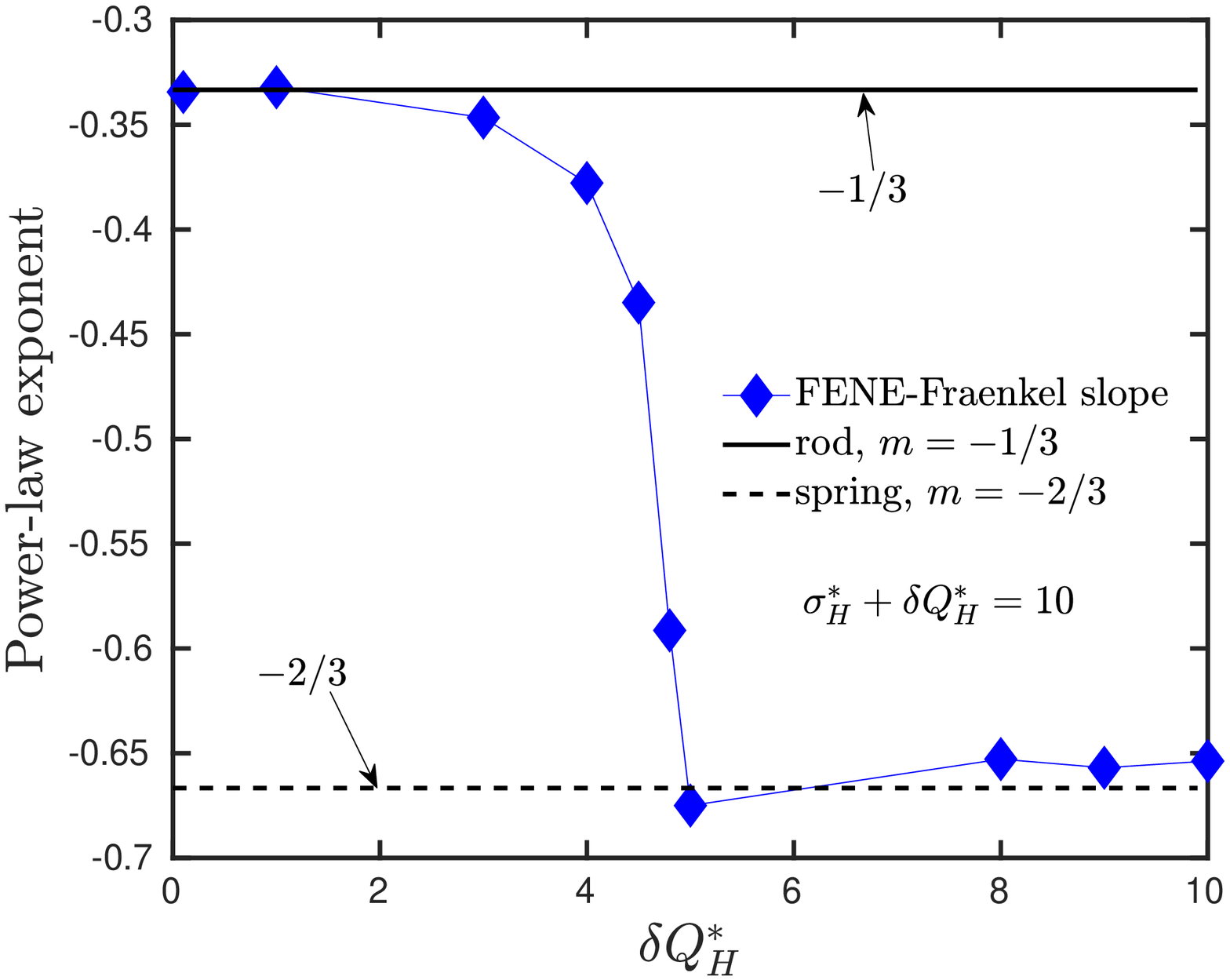} \\
       (b) \\
  \end{tabular}
  \caption[Crossover between rodlike FENE-Fraenkel spring and FENE spring]{Viscosity scaling with shear rate of FENE-Fraenkel dumbbells with a constant $\sigma^*_H + \delta Q^*_H = 10$ and $h^*_H = 0.15$. Fig.~(a) gives the viscosity curves for several $\delta Q^*_H$, while Fig.~(b) gives the high-shear power law exponent in the viscosity scaling at each $\delta Q^*_H$, determined by a linear fit to the last 3 data points. Lines are only guides for the eye and do not represent fits to the data. Note that quantities are scaled using Hookean units.}
  \label{Rod to FENE Viscosity}
\end{figure}

\begin{figure}[!ht]
  \centerline{
  \begin{tabular}{c}
        \includegraphics[width=8.5cm,height=!]{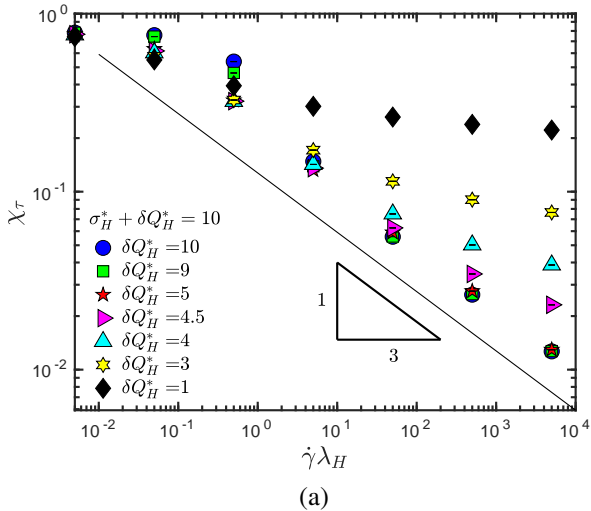} \\
        (a) \\
       \includegraphics[width=8.5cm,height=!]{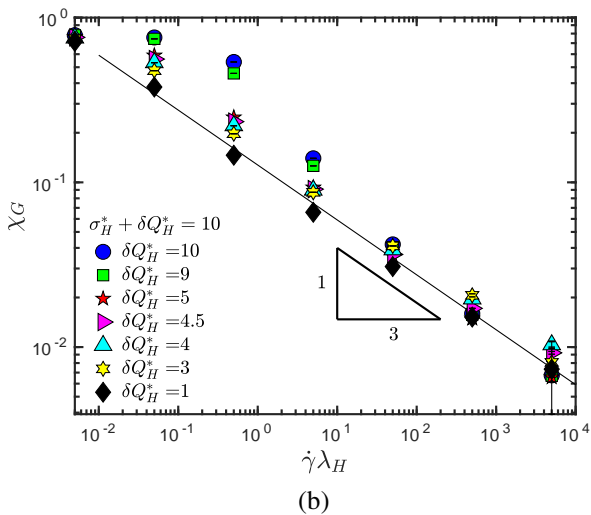} \\
       (b) \\
  \end{tabular}
  }
  \caption[$\chi_G$ and $\chi_{\tau}$ for rod to FENE]{Plot of $\chi_{\tau}$ (a) and $\chi_G$ (b) for the same parameter set as in Fig.~\ref{Rod to FENE Viscosity}. Lines are a guide to the eye and not fits to the data. Quantities are scaled using Hookean units.}
  \label{chi scaling rod to FENE}
  \vskip-10pt 
\end{figure}

\subsubsection{Rod to FENE-spring crossover in viscosity}
A FENE-Fraenkel spring with $\sigma = 0$ is identical to a FENE spring with the FENE $b$-parameter $b = {\delta Q^*_H}^2 \equiv H Q_0^2 / k_\mathrm{B} T$, where $Q_0$ is the maximum spring extensibility corresponding to $\delta Q$ for the FENE-Fraenkel spring when $\sigma = 0$.
At high shear rates, a FENE spring will show a $-2/3$ power-law scaling in the viscosity, while a rod (or a sufficiently stiff FENE-Fraenkel spring, as seen previously) will show a $-1/3$ scaling.
In order to compare these two regimes, Hookean non-dimensionalisation must be used, as there is no `FENE-limit' in the rodlike unit system.
A more extensible or less rodlike spring then corresponds to a lower $\sigma^*_H$ and a higher $\delta Q^*_H$, so that a natural way to compare a `rodlike' and `FENE-like' FENE-Fraenkel spring is to keep $\sigma^*_H + \delta Q^*_H$ constant.
This is shown in Fig.~\ref{Rod to FENE Viscosity}, where $\sigma^*_H + \delta Q^*_H = 10$, such that the $\sigma^*_H = 0$ case corresponds to a FENE spring with $b = 100$, and the $\sigma^*_H = 9.9$ case corresponds to $H^*_R =98.01$ and $\delta Q^*_R = 0.0101$.
Note that this does imply $\delta Q^*_R$ can be greater than 1, in which case $\delta Q^*_R$ can be thought of as a maximum fractional extension, but the spring clearly still cannot compress to $Q<0$. 
Physically, one could interpret this as a very rough model of a set of polymers with constant contour length when fully stretched, but different distributions of end-to-end distances.
The bead-rod case then corresponds to a delta-function peaked at $Q = \sigma$. 

Examining Fig.~\ref{Rod to FENE Viscosity}~(a), two major changes in the viscosity-shear rate curve are apparent as the spring is made more extensible (higher $\delta Q^*_H$ and hence lower $\sigma^*_H$). 
The first is a drop in the apparent zero-shear viscosity. 
This is easily explained by the decreased average end-to-end distance of the dumbbell as $\sigma^*_H$ is decreased, since the zero-shear viscosity is proportional to the equilibrium dumbbell length. 
The second change is in the high-shear behaviour, where the power-law slope of the viscosity curve goes from $-1/3$ to $-2/3$. 
Interestingly, at the shear rates investigated here, this crossover happens fairly suddenly at around $\delta Q^*_H = 5$ to $\delta Q^*_H = 4.5$, as the $\delta Q^*_H > 5$ viscosities converge at high shear rates.
This effect is seen clearly in Fig.~\ref{Rod to FENE Viscosity} (b), which shows that the crossover from $-2/3$ to $-1/3$ power law exponent begins suddenly around $\delta Q^*_H \approx \sigma^*_H \approx 5$.
The exponent then goes smoothly to the rodlike limit of $-1/3$ as $\delta Q^*_H$ decreases further.
Furthermore, this crossover also corresponds to a leveling off in $\chi_\tau$, as seen in Fig.~\ref{chi scaling rod to FENE}.
In other words, the value of $\delta Q^*_H$ at which the shear-thinning slope deviates from $-2/3$ (in this case, $\delta Q^*_H \approx 5$) is the same at which $\chi_\tau$ shows a high-shear plateau similar to that seen in the rodlike results of Fig.~\ref{chi scaling rodlike}. 

Finally, as expected, a spring with low $\sigma^*_H$ ($\sigma^*_H = 1$, $\delta Q^*_H = 9$) is approximately equivalent to a FENE spring of similar extensibility ($\sigma^*_H = 0$, $\delta Q^*_H = 10$). 
In other words, there is no discontinuity between the $\sigma \rightarrow 0$ and $\sigma = 0$ cases.

\begin{figure}[t]
  \centering
  \includegraphics[width=8.5cm,height=!]{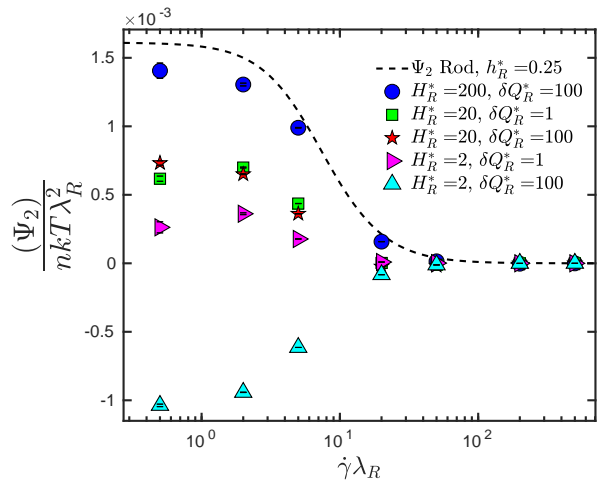}
  \caption[Second normal stress difference turns negative at low stiffness]{Second normal stress difference against shear rate for various spring extensibilities and stiffnesses at $h^*_R = 0.15$. Dotted line is $\Psi_2$ for the bead-rod dumbbell. This figure uses rodlike non-dimensionalisation. Error bars are smaller than symbol size.}
  \label{Psi2_negative_for_low_H_or_dQ}
  \vskip-10pt 
\end{figure}

\subsubsection{Changing sign of $\Psi_2$}
If fluctuations in hydrodynamic interactions are accounted for, either in BD simulations or using a Gaussian approximation, the second normal stress difference $\Psi_2$ of a Hookean dumbbell in shear flow will be strictly negative \cite{Ottinger1996, Prabhakar2002}. 
However, bead-rod dumbbells show a positive $\Psi_2$, so we should expect to see some sort of crossover as the extensibility is increased or the stiffness is decreased of the FENE-Fraenkel spring. 
Fig.~\ref{Psi2_negative_for_low_H_or_dQ} shows that this is in fact the case, where both a low stiffness and high extensibility is required to observe a negative $\Psi_2$.
In other words, only an analogue of a highly flexible polymer shows a negative $\Psi_2$, with more rigid models giving positive $\Psi_2$.
This figure again uses rodlike units to allow comparisons with bead-rod results.

This finding is significant given the history of theoretical and experimental determinations of $\Psi_2$ for high molecular weight polymers.
Up to about 1962, the `Weissenberg hypothesis' that $\Psi_2 = 0$ was thought to be correct, until several measurements showed that $\Psi_2$ may be positive \cite{bird1987dynamicsVol1, keentok1980measurement}.
This was then found to be due to subtle hole pressure effects, leading to a consensus that for flexible polymer solutions, $\Psi_2 < 0$.
This has been found in both rheometer force-based measurements using different plate geometries, as well as optical and shape-based measurements of channel flow \cite{alcoutlabi2009comparison, keentok1980measurement, ginn1969measurement}.
There seems to be a lack of similar measurements for rodlike molecules, with $\Psi_2$ changing sign with shear rate for highly concentrated solutions \cite{baek1993rheological}, but no clear results for dilute solutions.

Given the results of this computational study, it appears likely that a negative $\Psi_2$ is not a universal property of all polymer solutions, but instead a function of flexibility.
Further simulations of multi-bead chains with tunable flexibility may be necessary for investigation of this effect.
These results could be compared to experimental measurements of $\Psi_2$ for polymer solutions with a range of concentrations, flexibilities and morphologies (for example, using rigid macromolecules such as bacteriophages).

\begin{figure}[t]
  \centering
  \includegraphics[width=8.5cm,height=!]{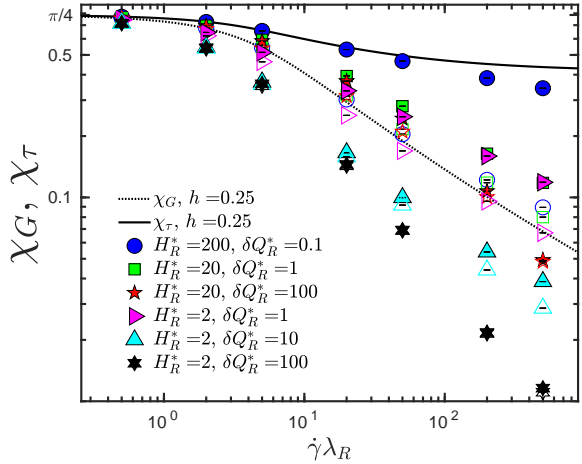}
  \caption[Influence of extensibility on $\chi_G$ and $\chi_{\tau}$ scaling]{Scaling of $\chi_G$ and $\chi_{\tau}$ with shear rate as extensibility and stiffness of the FENE-Fraenkel spring are varied at $h^*_R = 0.25$. Filled shapes are FENE-Fraenkel simulations of $\chi_{\tau}$, while hollow shapes are $\chi_G$. Rodlike units are used for comparison with rodlike results. Error bars are smaller than symbol size.}
  \label{chi extensibility}
 \vskip-15pt 
\end{figure}

\subsubsection{$\chi_G$ and $\chi_{\tau}$ scaling with extensibility}
\label{chi with extensibility}
The relative orientation of the gyration tensor (represented by $\chi_G$) and the stress tensor (represented by $\chi_{\tau}$) show complex behaviour depending on the stiffness and extensibility of the FENE-Fraenkel spring.
As Fig.~\ref{chi extensibility} shows, a low stiffness ($H^*_R = 2$) and highly extensible ($\delta Q^*_R = 100$) FENE-Fraenkel spring has similar $\chi_G$ and $\chi_{\tau}$ at all shear rates.
The slope of the $\chi_G$ and $\chi_{\tau}$ curves continue to decrease as $H$ is decreased or $\delta Q$ is increased, as the behaviour is fundamentally different in the Hookean limit.
As the extensibility is decreased, $\chi_G$ and $\chi_{\tau}$ move further apart, while increasing stiffness appears to change the absolute magnitude of $\chi_G$ and $\chi_{\tau}$ at a particular shear rate. 
This suggests that the FENE-Fraenkel spring captures behaviour which cannot be replicated individually by a FENE or Fraenkel spring.

As previously mentioned in section \ref{chi rods}, we should expect a high-shear plateau of the bead-rod $\chi_\tau$ at high shear rates due to the $-1/3$ power law scaling in $\eta$ and $-4/3$ power law scaling in $\Psi_1$.
For a FENE spring, the power law scaling in $\Psi_1$ remains at $-4/3$, while the viscosity now scales as $-2/3$, leading to an overall $-1/3$ scaling in $\chi_\tau$.
This can be seen in Fig.~\ref{chi scaling rod to FENE}~(a), where a more extensible spring leads to a power-law scaling in $\chi_\tau$.
Additionally, Fig.~\ref{chi extensibility} shows the same $-1/3$ scaling (similar to $\chi_G$) for FENE-Fraenkel dumbbells with low extensibility and low stiffness, but low stiffness and high extensibility (for example $H^*_R = 2$, $\delta Q^*_R = 100$, approaching a Hookean dumbbell) seems to lead to a further decrease in the power law exponent.

\subsection{Comparison with experimental data}
Here we show that the FENE-Fraenkel dumbbell model is able to give a reasonable description of the viscosity and $S$-parameter of semiflexible molecules with persistence length on the order of contour length.
We also compare results with rodlike models of the same aspect ratio and length as the experimentally measured molecules.
Although the FENE-Fraenkel spring is compared with real semiflexible polymers, the aim is not to develop this force law as a replacement for wormlike chain polymer models, or to suggest that the FENE-Fraenkel spring dumbbell can reproduce all the physics of these chains.
Instead, we wish to show that the FENE-Fraenkel dumbbell is at least as useful as a bead-rod dumbbell in modelling rigid polymers.
Additionally, the FENE-Fraenkel spring's adjustable extensibility may offer qualitative advantages over a rigid rod when the contour length is greater than the persistence length, since the variability in end-to-end distance of the true polymer chain may be somewhat captured by this extensiblity.

\subsubsection{Linear Dichroism Comparisons}
\label{experimental comparisons LD}

\begin{figure}[t]
  \centering
  \includegraphics[width=8.5cm,height=!]{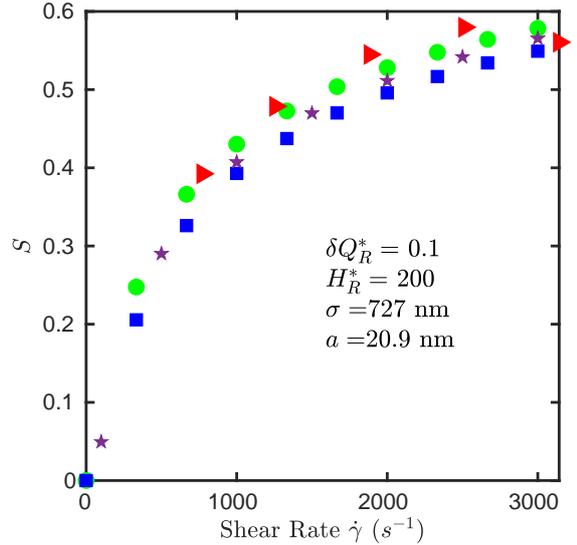}
  \caption[Rodlike model comparison with M13 Bacteriophage experimental LD]{Comparison of various rodlike models with experimental LD data for M13 bacteriophage with contour length $800$ nm and aspect ratio 100. Red triangles \tikzrighttriangle{red} are experimental LD measurements from \cite{McLachlan2013}, where the transition dipole moment was assumed to be $\alpha = 31^\circ$ to obtain the $S$-parameter values. Green circles \tikzcircle{green} are numerical calculations for an osculating multibead-rod model with bead diameter 8 nm \cite{bird1987dynamics}. Blue squares \tikzsquare{blue} are numerical calculations for a prolate spheroid with minor axis $4$ nm and major axis $400$ nm, identical to those of McLachlan et al. \cite{McLachlan2013}. Purple stars \tikzstar{magenta} are FENE-Fraenkel simulations with bead parameters displayed in the figure. Error bars are smaller than symbol size for the FENE-Fraenkel simulation, but were not provided for the experimental data.}
  \label{Rigid_model_S_comparisons}
%   \vskip-15pt 
\end{figure}

Fig.~\ref{Rigid_model_S_comparisons} compares the $S$-parameter prediction of three models (a FENE-Fraenkel spring dumbbell, a rigid multibead-rod and a prolate spheriod) with experimental data on the Linear Dichroism of M13 bacteriophage, which is filamentous with a persistence length ($\approx 1250$ nm \cite{Khalil2007}) longer than its contour length ($\approx 800$ nm) and a diameter of $\approx 8$ nm. 
The experimental data is taken from Ref. \cite{McLachlan2013}, where a microvolume Couette cell was used to shear M13 bacteriophage at several shear rates. 
As described in the paper from which this data was taken, the transition dipole moment angle for this bacteriophage can be determined via the protein structure of the phage to be $\alpha = 31^\circ$. 
Given the provided $LD_\text{r}$ data, the experimental $S$-parameter can be extracted and is plotted here. 
However, quantitative comparisons should be made cautiously, since experimental error and uncertainty in $\alpha$ were not specified.

The multibead-rod and prolate spheroid models can be applied without any fitting, given the aspect ratio and length of the original bacteriophage \cite{McLachlan2013, bird1987dynamics}.
However, the bead radius of the FENE-Fraenkel dummbell must be fit in some manner, as setting $a = 4$ nm gives completely inaccurate results.
This can be done by noting that for $S$-parameter prediction, the bead-rod dumbbell and multibead-rod are identical up to a factor proportional to the bead radius, such that the time constants characterising the two models can be equated using a simple analytical expression.
The bead radius $a = 20.9$ nm of the FENE-Fraenkel dumbbell was therefore chosen such that a bead-rod dumbbell with $a = 20.9$ nm would give exactly the same time constant and hence $S$-parameter prediction as the multibead-rod model with $a = 8$ nm.
A more detailed explanation of this procedure can be found in the supporing information section 5.
The values $H^*_R = 200$ and $\delta Q^*_R = 0.1$ were chosen based on the results of section \ref{material function results}, given that this set of parameters appears to fairly accurately reproduce a bead-rod dumbbell for moderate shear rates.
The value of $\sigma$ in dimensional form was then chosen such that $\sigma + \delta Q = 800$ nm, which gives $\sigma \approx 727$ nm. 

As can be seen in Fig.~\ref{Rigid_model_S_comparisons}, both the prolate spheroid and osculating multibead-rod, when given an aspect ratio equal to that of the bacteriophage, give results which appear quantitatively correct. 
This suggests that some form of slender rod is a suitable model for rigid macromolecules with persistence length close to or greater than contour length. 
This result is somewhat surprising given the persistence length of M13 is $\approx 1250$ nm \cite{Khalil2007} compared to its contour length of $800$ nm. 
Observed under a microscope, this molecule would appear quite bendy and flexible rather than perfectly rigid.
This provides an example of the suitability of a rigid segment in describing the behaviour of a polymer on length scales lower than the Kuhn length, while also suggesting that a segment model with some extensibility is no less physically reasonable than a perfectly rigid bead-bead link.
Specifically, the FENE-Fraenkel spring is able to predict the $S$-parameter, even though the extensibility is reasonably high ($10\%$ of the natural length $\sigma$). 
It seems that bead size and rodlike model parameters are more influential than the specific form of the force-extension curve or the connector model in general.
This shows that even without an extremely high $H$ or small $\delta Q$ to ensure strict matching to bead-rod results, the FENE-Fraenkel spring is a reasonable qualitative model of a fairly rigid molecule.

\subsubsection{Prediction of shear-dependent viscosity}
\begin{figure}[t]
  \centering
  \includegraphics[width=8.5cm,height=!]{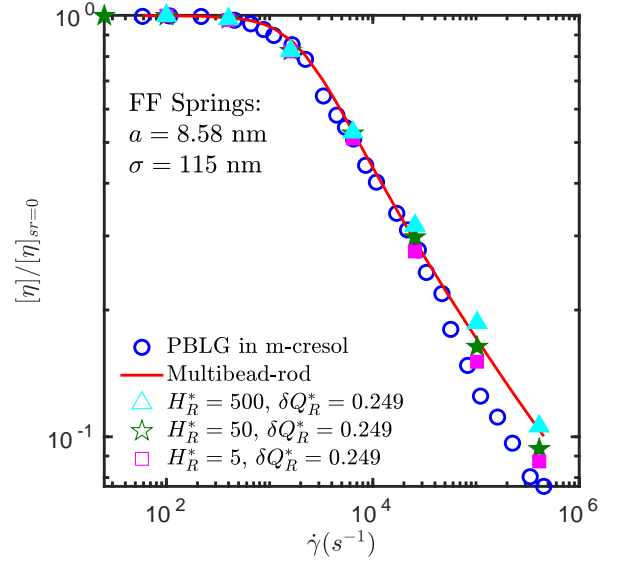}
  \caption[FENE-Fraenkel spring can reproduce viscosity curve for PBLG]{Comparison of experimental shear-rate dependent intrinsic viscosity $[\eta]$ for PBLG with a multibead-rod and FENE-Fraenkel spring. Experimental data is Poly-$\gamma$-benzyl-L-glutamate (PBLG) in m-cresol at $25.5^\circ$C, where the contour length is $L = 143$ nm and the persistence length is $90$ nm with an aspect ratio of $\approx 94$ \cite{yang1958non, Mead1990}. FENE-Fraenkel dumbbell bead radius was chosen to give a best fit to the experimental data.}
  \label{viscosity_comparison_PBLG}
  % \vskip-15pt 
\end{figure}

Yang \cite{yang1958non} has previously measured the shear-dependent intrinsic viscosity of Poly-$\gamma$-benzyl-L-glutamate (PBLG) in m-cresol solvent, which can be compared with our theoretical predictions.
This is displayed in Fig.~\ref{viscosity_comparison_PBLG}, which compares several FENE-Fraenkel spring dumbbells and a multibead-rod model with viscosity data for PBLG in m-cresol.
The FENE-Fraenkel spring parameters were chosen to roughly imitate a short wormlike chain with the same contour length and persistence length as PBLG (with contour length $L = 143$ nm and persistence length $l_p = 90 $ nm respectively \cite{yang1958non, Mead1990}).
The end-to-end distance distribution function for a short wormlike chain is given by Frey and Wilhelm \cite{wilhelm1996}, from which the equilibrium extension $\sqrt{\langle Q^2 \rangle_{\mathrm{eq}}}$ can be easily derived.
For the FENE-Fraenkel spring, $\sigma$ and $\delta Q$ were chosen such that $\sigma + \delta Q = L = 143$ nm and $\sigma = \sqrt{\langle Q_{\mathrm{WLC}}^2 \rangle_{\mathrm{eq}}}$, giving $\sigma = 115$ nm and $\delta Q = 28$ nm.
Finally, the bead radius $a$ was chosen in a similar way to that for the M13 bacteriophage comparisons in section \ref{experimental comparisons LD}, except that for viscosity there are two characteristic time constants for the bead-rod and multibead-rod models which cannot be directly equated. 
A bead-rod dumbbell radius of $a = 8.58$ nm was found to give results visually closest to a multibead-rod model with an aspect ratio of $\approx 94$ (the same aspect ratio as PBLG), so this bead radius was used for the FENE-Fraenkel spring dumbbell.
A more detailed explanation of this procedure can be found in the supporing information section 5.
Fig.~\ref{viscosity_comparison_PBLG} then shows results with three possible values of the spring stiffness $H^*_R$.

The stiffest spring, with $H^*_R = 500$, falls closest to the multibead-rod model, as would be expected for a more rodlike FENE-Fraenkel spring.
Interestingly, the springs with lower stiffness seem to predict the experimental results more accurately at high shear rates.
While it is difficult to draw solid conclusions from this single viscosity curve, it may be that the extensibility of the PBLG molecule is being revealed at higher shear rates, similar to the deviation of the extensible FENE-Fraenkel spring from rodlike viscosity at high shear rates.
In this way, a `rodlike' model which has finite extensibility, such as the FENE-Fraenkel spring, may be more useful than a true rigid rod when predicting shear viscosity of certain polymers.
Therefore, the FENE-Fraenkel spring is a promising force law with which to investigate the differences in high shear-rate behaviour of bead-rod and bead-spring chains.
Further research on FENE-Fraenkel spring chains may provide insight into why different models are only able to predict some behaviours of certain true polymer chains and not others.

%%%%%%%%%%%%%%%%%%%%%%%%%%%%%%%%%%%%%%%%%%%%%%%%%%%%%%%%%%%%%%%%%%%%%%%%%%%%%%%%%%%%%%%%%%%%%%%%%%%%%%%%%%%%%%%%%%%%%%%%%%%%%%%%%%%%%%%%%%%%%%%%%%%%%%%%%%%%%%%%%%%%%%%%%%%%%%%%%%%%%%%%%%%%%%%%%%%%%%%%

\section{Conclusions}
The FENE-Fraenkel spring has been shown to be a viable replacement for a rod in terms of the stress jump, material functions and rheo-optical properties.
However, there is no set of spring parameters which give universal adherence to bead-rod results.
Instead, the particular measured property and the shear rate also affect the calculated accuracy, such that stiffer and less extensible springs are needed for higher shear rates and parameters such as $\Psi_1$ and $\Psi_2$ as opposed to $\eta$ and $\chi$.
For example, using $H^*_R = 200$ and an extensibility of $\delta Q^*_R = 0.1$ is sufficient to obtain $\eta_\mathrm{rod}$ to within 1\% accuracy up to $\dot{\gamma}^*_R = 50$ (Fig.~\ref{FF_Rigid_comparison_eta}), while a far stiffer $H^*_R = 5000$ and $\delta Q^*_R = 0.02$ is required to obtain $\Psi_{2, \mathrm{rod}}$ to within 10\% for $\dot{\gamma}^*_R > 50$ (Fig.~\ref{FF_Rigid_comparison_Psi2}).

There is also the issue of timestep convergence, which should be checked explicitly for each $h^*$, $\delta Q^*_H$ and $\dot{\gamma}$, which were found to be the key factors influencing the required timestep for convergence of the equilibrium distribution function and non-equilibrium averages.
In general, as $\delta Q^*_H$ is lowered a smaller $\Delta t^*_H$ is needed for distribution function convergence at equilibrium, as seen in Fig.~\ref{equilibrium timestep convergence with dQ}.
This gives a good starting point to check timestep convergence at non-zero shear rates.

As the spring stiffness and extensibility are relaxed towards the FENE, Fraenkel and Hookean limits, the FENE-Fraenkel spring begins to show more traditionally `spring-like' behaviour.
This manifests itself most clearly in a change in the shear-thinning exponent of the viscosity, which moves from $-1/3$ to $-2/3$ as $H$ is decreased and $\delta Q$ is increased.
It appears that any FENE-Fraenkel spring will eventually deviate from the $-1/3$ scaling at sufficiently high shear rates, with a higher stiffness or lower extensibility causing this change in scaling to occur at higher and higher $\dot{\gamma}$.
Therefore, higher shear rates appear to reveal more of the `spring-like' nature of a stiff and inextensible FENE-Fraenkel spring.
Additionally, the second normal stress difference appears negative for the FENE-Fraenkel spring only when both $H^*_R$ is small and $\delta Q^*_R$ is large, converging towards Hookean behaviour.
Significantly, the FENE-Fraenkel spring is able to represent both positive and negative $\Psi_2$ with a single form of the spring potential.
This is also the case for $\chi_\tau$ and $\chi_G$, where the stress optical law does not hold if either $H^*_R$ is large or $\delta Q^*_R$ is small.
Once again this demonstrates that the FENE-Fraenkel spring is able to represent both spring and rod regimes with correctly chosen parameters.

Comparisons with experimental data show the FENE-Fraenkel spring is also able to qualitatively reproduce the behaviour of rigid filamentous molecules.
With spring parameters of $H^*_R = 200$ and $\delta Q^*_R = 0.1$, which were sufficient to reproduce bead-rod behaviour at low shear rates, the FENE-Fraenkel spring can accurately model the linear dichroism of M13 bacteriophage, given an appropriately chosen bead radius $a$.
Of particular note is data on the shear-thinning of PBLG polymer in m-cresol solvent, which displays a slight deviation from the classic $-1/3$ power law exponent at very high shear rates.
This may be due to the fact that this polymer is not truly rigid and has some flexibility and extensibility, which the FENE-Fraenkel spring is able to qualitatively capture.
This suggests that the deviation of the FENE-Fraenkel spring from bead-rod material properties at high shear rates may not be an issue when used as a model for links in a semi-flexible polymer, as experimental measurements also display this behaviour.

Future work will focus on two main areas.
Firstly, whether the results obtained for the dumbbell case when comparing a FENE-Fraenkel spring and rod remain valid in the bead-spring-chain case.
Secondly, whether the FENE-Fraenkel spring can be used, along with a bending potential, as a model of a semiflexible polymer chain such as DNA in the context of modelling the Linear Dichroism signal in shear flow.
In this second case, the ability of the FENE-Fraenkel spring to represent both a rod and an entropic spring may be useful when comparing different possible levels of coarse-graining to use in the model.

\appendix



%% References



%% New version of the num-names style
\bibliographystyle{elsarticle-num-names}
\bibliography{JNNFM-S-20-00216R1_arXiV.bib}






% \end{thebibliography}

\end{document}